\documentclass[zpreprint,zbstepj]{./LaTeX/zeus/zeus_paper}
\usepackage[english]{babel}
\chardef\usc=95
\chardef\til=126
\catcode`\@=11 
\DeclareRobustCommand\xdotspace{\futurelet\@let@token\@xdotspace}
\def\@xdotspace{%
  \ifx\@let@token.\else
  \ifx\@let@token\bgroup.\else
  \ifx\@let@token\egroup.\else
  \ifx\@let@token\/.\else
  \ifx\@let@token\ .\else
  \ifx\@let@token~.\else
  \ifx\@let@token!.\else
  \ifx\@let@token,.\else
  \ifx\@let@token:.\else
  \ifx\@let@token;.\else
  \ifx\@let@token?.\else
  \ifx\@let@token/.\else
  \ifx\@let@token'.\else
  \ifx\@let@token).\else
  \ifx\@let@token-.\else
  \ifx\@let@token\@xobeysp.\else
  \ifx\@let@token\space.\else
  \ifx\@let@token\@sptoken.\else
   .\space
   \fi\fi\fi\fi\fi\fi\fi\fi\fi\fi\fi\fi\fi\fi\fi\fi\fi\fi}
\catcode`\@=12 

\newcommand{\stru}[2]{%
   \relax\ifmmode\hbox{\vrule height#1 depth#2 width0pt}%
   \else\vrule height#1 depth#2 width0pt\fi}

\newcommand{\Ronum}[1]{\uppercase\expandafter{\romannumeral#1}}
\newcommand{\ronum}[1]{\expandafter{\romannumeral#1}}
\DeclareRobustCommand{\LaTeXZ}{%
  \LaTeX\kern-.05em4\kern-.1em
  {\raisebox{-0.2ex}{$\scriptstyle\text{ZEUS}$}}\xspace}

\DeclareMathAlphabet{\mathbf}{OT1}{cmr}{bx}{sl}
\newcommand{\eVdist}{\kern-0.06667em}

\newcommand{\slashfrac}[2]{%
  \raisebox{0.5ex}{\ensuremath #1}\kern-0.12em/\kern-0.08em
  \raisebox{-.8ex}{\ensuremath #2}}

\newcommand{\sqr}[3]{%
    {\vcenter{\hrule height.#3ex\hbox{\vrule width.#2ex height#1ex
     \kern#1ex\vrule width.#3ex}\hrule height.#2ex}}}

\catcode`\@=11 
\newcommand{\parenbar}{\mathpalette\p@renb@r}
\def\p@renb@r#1#2{\vbox{%
  \ifx#1\scriptscriptstyle \dimen@.7em\dimen@ii.2em\else
  \ifx#1\scriptstyle \dimen@.8em\dimen@ii.25em\else
  \dimen@1em\dimen@ii.4em\fi\fi \offinterlineskip
  \ialign{\hfill##\hfill\cr
    \vbox{\hrule width\dimen@ii}\cr
    \noalign{\vskip-.3ex}%
    \hbox to\dimen@{$\mathchar300\hfil\mathchar301$}\cr
    \noalign{\vskip-.3ex}%
    $#1#2$\cr}}}
\catcode`\@=12 

\newcommand{\IP}{{\rm I$\kern-0.01667em$P}\xspace}

\mathchardef\qsm=63
\mathchardef\pls=43
\mathchardef\mns=512
\mathchardef\plm=518
\mathchardef\eql=61
\mathchardef\smallleft=300
\mathchardef\smallright=301
\mathchardef\les=316
\mathchardef\gre=318
\mathchardef\leq=532
\mathchardef\grq=533

\catcode`\@=11 
\newcounter{pict@width}
\newcounter{pict@height}
\newlength{\pict@scale}
\newcommand{\psfigadd}[4]{%
\setcounter{pict@width}{1*\ratio{#2+\pict@scale/2}{\pict@scale}}
\setcounter{pict@height}{1*\ratio{#3+\pict@scale/2}{\pict@scale}}
\setlength{\unitlength}{\pict@scale}
\hbox to #2{\hspace{-\fill}\begin{picture}(\thepict@width,\thepict@height)
\put(0,0){\psfig{figure=#1,width=#2,height=#3,clip=}}
\SetScale{0.283466457}
\SetWidth{1.763889}
{#4}
\end{picture}}
}
\newcounter{pict@widthfst}
\newcounter{pict@widthscd}
\newcounter{pict@widthtot}
\newcommand{\psfigaddtwo}[7]{%
\setcounter{pict@widthfst}{1*\ratio{#2+\pict@scale/2}{\pict@scale}}
\setcounter{pict@widthscd}{1*\ratio{#2+#4+\pict@scale/2}{\pict@scale}}
\setcounter{pict@widthtot}{1*\ratio{#2+#4+#6+\pict@scale/2}{\pict@scale}}
\setcounter{pict@height}{1*\ratio{#3+\pict@scale/2}{\pict@scale}}
\setlength{\unitlength}{\pict@scale}
\hbox{\hspace{-\fill}\begin{picture}(\thepict@widthtot,\thepict@height)
\put(0,0){\psfig{figure=#1,width=#2,height=#3,clip=}}
\put(\thepict@widthscd,0){\psfig{figure=#5,width=#6,height=#3,clip=}}
\SetScale{0.283466457}
\SetWidth{1.763889}
{#7}
\end{picture}}
}
\newcommand{\psfigror}[4]{%
\setcounter{pict@width}{1*\ratio{#2+\pict@scale/2}{\pict@scale}}
\setcounter{pict@height}{1*\ratio{#3+\pict@scale/2}{\pict@scale}}
\setlength{\unitlength}{\pict@scale}
\hbox{\begin{picture}(\thepict@width,\thepict@height)
\put(0,\thepict@height){\psfig{figure=#1,width=#3,height=#2,clip=,angle=270}}
\SetScale{0.283466457}
\SetWidth{1.763889}
{#4}
\end{picture}}
}
\newcommand{\psfigrol}[4]{%
\setcounter{pict@width}{1*\ratio{#2+\pict@scale/2}{\pict@scale}}
\setcounter{pict@height}{1*\ratio{#3+\pict@scale/2}{\pict@scale}}
\setlength{\unitlength}{\pict@scale}
\hbox{\begin{picture}(\thepict@width,\thepict@height)
\put(0,0){\psfig{figure=#1,width=#3,height=#2,clip=,angle=90}}
\SetScale{0.283466457}
\SetWidth{1.763889}
{#4}
\end{picture}}
}
\catcode`\@=12 
\newlength\listtextwidth

\catcode`\@=11 
\newlength{\@tabfninsert}
\newlength{\@tabfnwidth}
\newcommand{\tabfootnote}[2]{%
  \setlength{\@tabfninsert}{0.8em}
  \setlength{\@tabfnwidth}{\textwidth}
  \addtolength{\@tabfnwidth}{-\@tabfninsert}
  \addtolength{\@tabfnwidth}{-0.4em}
  \noindent\makebox[\@tabfninsert][r]{\footnotesize$^{#1}$\hfil}\hfill%
  \parbox[t]{\@tabfnwidth}{\footnotesize #2\hfill}}
\catcode`\@=12 

\newcommand{\xgo} {\mbox{$x_{\gamma}^{\rm obs}$}}
\newcommand{\cts} {\mbox{$\cos\theta^*$}}

\def\3{\ss}

\def\citeF2G{{\cite{%
zfp:c24:231,zfp:c31:527,np:b281:365,zfp:c34:1,epj:c18:15,*pl:b411:387,*zfp:c61:199,%
*zfp:c74:33,*pl:b412:225,zfp:c69:223,pl:b436:403,*pl:b447:147%
}}\xspace}

\begin{document}
\title{
\vspace{-5cm}
\begin{flushleft} {\normalsize \tt DESY 01-220}\\ \vspace{-.25cm}{\normalsize \tt December 2001} \end{flushleft}
\vspace{2cm}
Dijet photoproduction at HERA and the structure of the photon
}                                                       
                    
\author{ZEUS Collaboration}


\abstract{
The dijet cross section in photoproduction has been measured with the ZEUS detector at HERA 
using an integrated luminosity of 38.6~pb$^{-1}$. The events were required to have a 
virtuality of the incoming photon, $Q^2$, of less than 1~GeV$^2$ and a photon-proton 
centre-of-mass energy in the range $134 < W_{\gamma p} < 277$~GeV. Each event contains at 
least two jets satisfying transverse-energy requirements of $E_{T}^{\rm jet1}>14$ GeV and  
$E_{T}^{\rm jet2}>11$ GeV and pseudorapidity requirements of \mbox{$-1<\eta^{\rm jet1,2}<2.4$}. 
The measurements are compared to next-to-leading-order QCD predictions.
The data show particular sensitivity to the density of partons in the photon, allowing the 
validity of the current parameterisations to be tested.}

\makezeustitle

\def\3{\ss}                                                                                        
\pagenumbering{Roman}                                                                              
                                                   %
\begin{center}                                                                                     
{                      \Large  The ZEUS Collaboration              }                               
\end{center}                                                                                       
  S.~Chekanov,                                                                                     
  M.~Derrick,                                                                                      
  D.~Krakauer,                                                                                     
  S.~Magill,                                                                                       
  B.~Musgrave,                                                                                     
  A.~Pellegrino,                                                                                   
  J.~Repond,                                                                                       
  R.~Yoshida\\                                                                                     
 {\it Argonne National Laboratory, Argonne, Illinois 60439-4815}~$^{n}$                            
\par \filbreak                                                                                     
  M.C.K.~Mattingly \\                                                                              
 {\it Andrews University, Berrien Springs, Michigan 49104-0380}                                    
\par \filbreak                                                                                     
  P.~Antonioli,                                                                                    
  G.~Bari,                                                                                         
  M.~Basile,                                                                                       
  L.~Bellagamba,                                                                                   
  D.~Boscherini,                                                                                   
  A.~Bruni,                                                                                        
  G.~Bruni,                                                                                        
  G.~Cara~Romeo,                                                                                   
  L.~Cifarelli,                                                                                    
  F.~Cindolo,                                                                                      
  A.~Contin,                                                                                       
  M.~Corradi,                                                                                      
  S.~De~Pasquale,                                                                                  
  P.~Giusti,                                                                                       
  G.~Iacobucci,                                                                                    
  G.~Levi,                                                                                         
  A.~Margotti,                                                                                     
  T.~Massam,                                                                                       
  R.~Nania,                                                                                        
  F.~Palmonari,                                                                                    
  A.~Pesci,                                                                                        
  G.~Sartorelli,                                                                                   
  A.~Zichichi  \\                                                                                  
  {\it University and INFN Bologna, Bologna, Italy}~$^{e}$
% ...
\newpage                                                                                           
$^{\    1}$ now at Cornell University, Ithaca/NY, USA \\                                              
$^{\    2}$ on leave of absence at University of                                                   
Erlangen-N\"urnberg, Germany\\                                                                     
% ...
                                                           %
                                                           %
\newpage   
                                                           %
                                                           %
\begin{tabular}[h]{rp{14cm}}                                                                       
$^{a}$ &  supported by the Natural Sciences and Engineering Research                               
          Council of Canada (NSERC) \\                                                             
$^{b}$ &  supported by the German Federal Ministry for Education and                               
          Research (BMBF), under contract numbers HZ1GUA 2, HZ1GUB 0, HZ1PDA 5, HZ1VFA 5\\         
$^{c}$ &  supported by the MINERVA Gesellschaft f\"ur Forschung GmbH, the                          
          Israel Science Foundation, the U.S.-Israel Binational Science                            
          Foundation, the Israel Ministry of Science and the Benozyio Center                       
          for High Energy Physics\\                                                                
$^{d}$ &  supported by the German-Israeli Foundation, the Israel Science                           
          Foundation, and by the Israel Ministry of Science\\                                      
$^{e}$ &  supported by the Italian National Institute for Nuclear Physics (INFN) \\                
$^{f}$ &  supported by the Japanese Ministry of Education, Science and                             
          Culture (the Monbusho) and its grants for Scientific Research\\                          
$^{g}$ &  supported by the Korean Ministry of Education and Korea Science                          
          and Engineering Foundation\\                                                             
$^{h}$ &  supported by the Netherlands Foundation for Research on Matter (FOM)\\                   
$^{i}$ &  supported by the Polish State Committee for Scientific Research,                         
          grant no. 115/E-343/SPUB-M/DESY/P-03/DZ 121/2001-2002\\                                  
$^{j}$ &  partially supported by the German Federal Ministry for Education                         
          and Research (BMBF)\\                                                                    
$^{k}$ &  supported by the Fund for Fundamental Research of Russian Ministry                       
          for Science and Edu\-cation and by the German Federal Ministry for                       
          Education and Research (BMBF)\\                                                          
$^{l}$ &  supported by the Spanish Ministry of Education and Science                               
          through funds provided by CICYT\\                                                        
$^{m}$ &  supported by the Particle Physics and Astronomy Research Council, UK\\                   
$^{n}$ &  supported by the US Department of Energy\\                                               
$^{o}$ &  supported by the US National Science Foundation                                          
\end{tabular}                                                                                      
                                                           %
                                                           %


\pagenumbering{arabic} 
\pagestyle{plain}
\section{Introduction}
\label{sec-int}

\begin{figure}[htb]
\begin{center}
~\epsfig{file=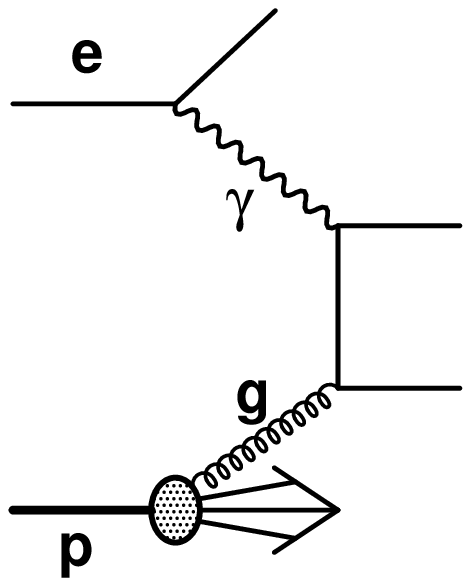}
\hspace{1cm}~\epsfig{file=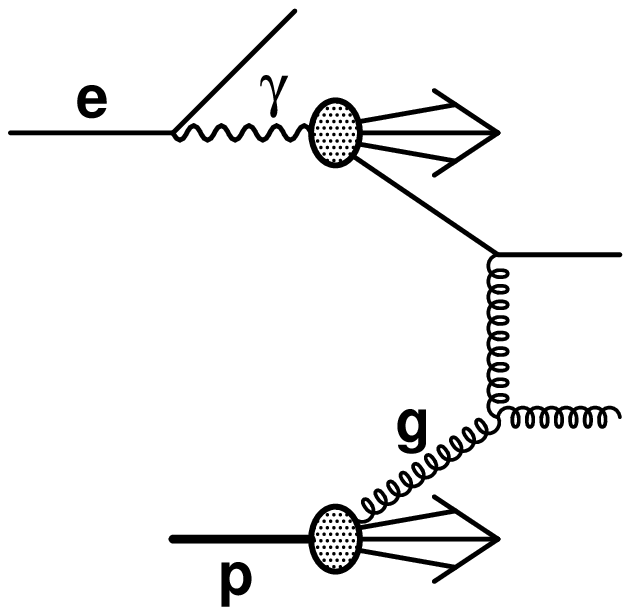}
\put(-305,6){\makebox(0,0)[tl]{\large (a)}}  
\put(-95,6){\makebox(0,0)[tl]{\large (b)}}  
\end{center}
\caption{Examples of (a) direct and (b) resolved dijet photoproduction diagrams in LO QCD.}
\label{fig:feyn}
\end{figure}

In photoproduction at HERA, a quasi-real photon, emitted from the
incoming positron, collides with a parton from the incoming proton.
The photoproduction of jets can be classified into two types of process
in leading-order (LO) Quantum Chromodynamics (QCD).
In direct processes (Fig.~\ref{fig:feyn}(a)), the photon participates
in the hard scatter via either boson-gluon fusion or QCD Compton
scattering. The second class, resolved processes
(Fig.~\ref{fig:feyn}(b)), involve the photon acting as a source of
quarks and gluons, with only a fraction of its momentum participating
in the hard scatter.  Measurements of jet cross sections in
photoproduction~\cite{pl:b483:36,*epj:c1:97,*zfp:c70:17,*pl:b314:436,epj:c11:35,pl:b443:394,epj:c1:109,pl:b384:401,pl:b348:665,pl:b322:287,epj:c18:625,*epj:c6:67,*epj:c4:591,*pl:b342:417} 
are thus sensitive to the structure of the photon and the proton, and to 
the dynamics of the hard sub-processes as calculated in
perturbative QCD (pQCD). These jet cross sections can
therefore be used in global fits to data to determine the
parton densities in both the photon and proton.

In the kinematic range of the measurements presented in this paper, the value of $x_p$, the 
fractional momentum at which partons inside the proton are probed, lies predominantly in the 
region between $0.01$ and $0.1$. At these $x_p$ values, the parton densities in the proton are 
constrained by measurements of the structure function, 
$F_2^p$,~\cite{pl:b223:485,*np:b483:3,*epj:c21:33,*epj:c21:443} in deep inelastic scattering 
(DIS). The present measurements are directly sensitive to both the quark and the gluon content 
of the photon. The fractional momentum of the photon carried by the interacting parton, 
$x_\gamma$, lies between 0.1 and 1. For $x_\gamma$ values above $0.5$, the quark densities in 
the photon are not well constrained by $F_2^\gamma$ data obtained from $\gamma \gamma^*$ 
scattering in $e^+e^-$ experiments~\citeF2G. The gluon density, to which jet photoproduction 
is directly sensitive at LO, is also poorly constrained by the $F_2^\gamma$ data for all 
$x_\gamma$. The most recent measurements of $F_2^\gamma$ from LEP extend up to an average 
scale of $\sim 25$~GeV. These, and higher scales, can be studied in jet production at 
HERA.

The aim of the present investigation is to provide constraints, from data on dijet 
photoproduction, on the parton densities in the photon in the range \mbox{$0.1~<~x_\gamma~<~1$} 
and to probe the dynamics of the hard sub-processes. For this purpose, the dijet cross section 
is measured at high transverse energies where next-to-leading-order (NLO) QCD calculations are 
expected to describe the data. In this kinematic region, where the effects of soft physics are  
suppressed and the parton densities in the proton are well known, the data can be used to test 
the validity of the current parameterisations of the parton densities in the photon. At high 
$x_\gamma$, where the effects of the uncertainties in the photon structure are small, these 
data also provide a consistency check of the gluon distribution in the proton extracted from 
deep inelastic scattering. 

This analysis builds on the improved understanding of jet photoproduction and of comparisons to 
NLO QCD calculations gained in previous analyses~\cite{epj:c1:109,epj:c11:35}. With an increase 
of a factor of six in luminosity, an extension of the kinematic region and reduced systematic 
uncertainties, the measurements in this paper\footnote{After submission of this paper, 
a similar study from the H1 collaboration became available~\cite{desy-01-225}.} have greatly 
improved precision compared to the previous ZEUS dijet measurement~\cite{epj:c11:35}.

\section{Theoretical framework}
\label{sec:frame}

Within the framework of pQCD, the dijet photon-proton cross section, $d\sigma_{\gamma p}$, can be 
written as a convolution of the proton parton density functions (PDFs), $f_p$, and photon PDFs, 
$f_\gamma$, with the partonic hard cross section, $d\hat{\sigma}_{ab}$:
  
\[ 
d\sigma_{\gamma p} = \sum_{ab} \int_{x_p} \int_{x_\gamma} dx_p dx_\gamma f_p(x_p, \mu^2) 
f_\gamma(x_\gamma, \mu^2) d\hat{\sigma}\mathit{_{ab}} (x_p, x_\gamma, \mu^2)
\cdot (1+\delta_{\rm had}).
\]

For photoproduction cross sections measured in lepton-proton scattering, there is an additional 
convolution with the distribution of photons from the lepton beam. In the case of the direct 
cross section, the photon structure is replaced by a delta function at $x_\gamma=1$. The scale 
of the process, $\mu$, represents both the renormalisation scale, $\mu_R$, and factorisation 
scale, $\mu_F$, which are set equal for this study. The hadronisation correction, 
$\delta_{\rm had}$, accounts for 
non-perturbative effects in the final state and can be estimated using Monte Carlo (MC) models for 
the parton cascade and fragmentation; it is, in general, a function of the variable being measured
(see Section~\ref{sect:mc}).

The distribution of the dijet angle, $\theta^*$, in the parton-parton centre-of-mass frame 
is directly sensitive to the form of the matrix elements, and hence to 
the partonic hard cross section. For massless partons, the centre-of-mass scattering angle 
is given by

\[ \cts = \tanh \left( \frac{\eta^{\rm jet1}-\eta^{\rm jet2}}{2} \right) , \]

where $\eta^{\rm jet1}$ and $\eta^{\rm jet2}$ are the pseudorapidities in the laboratory 
frame of the two jets of highest transverse energy. Only the absolute value of $\cts$ can be 
determined because the originating parton cannot be identified. The variable $\cts$ is 
invariant under the different boosts along the beam axis arising from the spectrum of incoming 
parton momenta. This minimises the sensitivity of the differential cross section, $d\sigma/d\cts$, 
to the momentum density distribution of the partons in the photon and proton.

For jets of transverse energy of more than 6 GeV, it has been shown~\cite{pl:b384:401} that samples 
of events enriched in either direct or resolved photon processes have very different angular 
distributions. The cross section for the sample enriched in resolved photon events increases more 
rapidly at high $\cts$ than that in direct photon events. This is expected at both LO and NLO 
QCD~\cite{pr:d40:2844}; both predictions give a good description of the data. The different angular 
dependence of the cross sections can be explained in terms of the dominant propagators in the 
respective samples.
In direct events, the dominant processes (mostly boson-gluon fusion) have a spin--$\frac{1}{2}$ 
quark propagator and the angular dependence is approxiamtely $\propto (1-|\cts|)^{-1}$. 
In resolved events (e.g. $qg \rightarrow qg$ and $gg \rightarrow gg$), the dominant processes 
have a spin--1 gluon propagator, which has an angular dependence $\propto (1-|\cts|)^{-2}$.

To probe the structure of the photon, it is desirable to measure cross sections as functions of 
variables that are sensitive to the spectrum of incoming parton momenta, such as the pseudorapidity 
of the jets or the momentum fraction, $x_\gamma$. Since $x_\gamma$ is not directly measurable, an 
observable, \xgo, which is the fraction of the photon's momentum participating in the production of 
the two highest transverse-energy jets (and is equal to $x_\gamma$ for partons in LO QCD), is 
introduced~\cite{pl:b348:665}:  

\[ \xgo = \frac{E^{\rm jet1}_{T} e^{-\eta^{\rm jet1}} 
           + E^{\rm jet2}_{T} e^{-\eta^{\rm jet2}}}
       {2yE_e}. \]

Here $E^{\rm jet1}_{T}$ and $E^{\rm jet2}_{T}$ are the transverse energies of the two jets in the 
laboratory frame ($E^{\rm jet1}_{T} > E^{\rm jet2}_{T}$) and $y$ is the fraction of the 
positron's energy, $E_e$, carried by the photon in the proton rest frame. The quantity \xgo \ is a 
particularly useful variable with which to discriminate between various photon PDFs.

\section{Experimental conditions}
\label{sec-exp}

The data were collected during the 1996 and 1997 running periods, when 
HERA operated with protons of energy $E_p~=~820~{\mbox{GeV}}$ and positrons of energy 
$E_e~=~27.5$~GeV, and correspond to an integrated luminosity of 
$38.6\pm0.6~\mbox{pb}^{-1}$. A detailed description of the ZEUS detector can be found 
elsewhere~\cite{pl:b297:404,zeus:1993:bluebook}. A brief outline of the components that are most 
relevant for this analysis is given below.

The high-resolution uranium-scintillator calorimeter 
(CAL)~\cite{nim:a309:77,*nim:a309:101,*nim:a336:23,*nim:a321:356} 
consists of three parts: the forward, the 
barrel and the rear calorimeters. Each part is subdivided transversely into towers and 
longitudinally into one electromagnetic and either one (in the rear) or two (in the barrel and 
forward) 
hadronic sections. The smallest subdivision of the calorimeter is called a cell. The CAL 
relative energy resolutions, as measured under test-beam conditions, are $0.18/\sqrt{E}$ for 
electrons and $0.35/\sqrt{E}$ for hadrons ($E$ in GeV). 

Charged particles are measured in the central tracking detector 
(CTD)~\cite{nim:a279:290,*npps:b32:181,*nim:a338:254}, which operates in a magnetic field of 1.43~T 
provided by a thin superconducting coil. The CTD covers the polar-angle\footnote{The ZEUS 
coordinate system is a right-handed Cartesian system, with the $Z$ axis pointing in the proton 
beam direction, referred to as the ``forward direction'', and the $X$ axis pointing left towards 
the centre of HERA. The coordinate origin is at the nominal interaction point. The 
pseudorapidity is defined as $\eta = -\ln(\tan \frac{\theta}{2})$, where the polar angle, 
$\theta$, is measured with respect to the proton beam direction.} region 
\mbox{$15^\circ<\theta<164^\circ$}. The relative transverse-momentum 
resolution for full-length tracks is $\sigma_{p_T}/p_T~=~0.0058p_T \oplus 0.0065 \oplus 0.0014/p_T$ 
with $p_T$ in GeV. Tracking information along with energy deposits in the CAL were used to measure 
the transverse energy and direction of jets as described in detail in Section~\ref{energycorr}.

The luminosity was measured from the rate of the bremsstrahlung process 
$e^+p~\rightarrow~e^+\gamma p$, where the photon was measured in a lead-scintillator 
calorimeter~\cite{desy-92-066,*zfp:c63:391,*desy-01-141} placed in the HERA tunnel at $Z=-107~{\rm m}$.

A three-level trigger system was used to select events online~\cite{zeus:1993:bluebook,epj:c1:109}. 
At the third level, a cone algorithm was applied to the CAL cells and jets were reconstructed using 
the energies and positions of these cells. Events with at least two jets, each of which 
satisfied the requirements that the transverse energy exceeded 4 GeV and the pseudorapidity was 
less than 2.5, were accepted.

\section{Definition of the cross section}
\label{photon_structure}

The kinematic region for this study is the photoproduction region, defined as $Q^2<1 {\rm GeV^2}$, 
with a photon-proton centre-of-mass energy, $W_{\gamma p}$, in the range 134~GeV to 277~GeV. Each 
event is required to have at least two jets reconstructed with the $k_T$ cluster 
algorithm~\cite{np:b406:187} in its longitudinally invariant inclusive mode~\cite{pr:d48:3160}, 
with at least one 
jet having transverse energy greater than 14~GeV and another greater than 11~GeV. The jets are 
required to satisfy $-1<\eta^{\rm jet1,2}<2.4$, an extension of the pseudorapidity 
range by 0.4 units in the forward direction over the previous analysis~\cite{epj:c11:35}, thereby 
increasing the sensitivity of the measurement to resolved photon processes.

Cross sections are presented as a function of \xgo, $E_{T}^{\rm jet1}$ and $\eta^{\rm jet2}$. 
The cross sections for jet variables are symmetrised~\cite{prl:73:2019} with respect to the 
pseudorapidities of the 
two jets. Each event, therefore, contributes twice to the cross section. The 
cross sections have been determined for regions enriched in direct and resolved photon processes by 
requiring  \xgo \ to be greater than 0.75 or less than 0.75, respectively.

Additional kinematic constraints were applied to the measurement of the cross section as a function 
of $|\cts|$ to remove biases imposed by the other requirements. For a given centre-of-mass energy, 
events at high $|\cts|$ have small scattering angles and thus lower $E_T^{\rm jet}$. To study the 
$|\cts|$ distribution up to $|\cts| = 0.8$ without bias from the $E_T^{\rm jet}$ requirements, a 
cut on the dijet mass of $M^{\rm jj} > 42$~GeV was applied. The dijet mass is defined in terms of 
the two jets with highest transverse energy as

\[ M^{\rm jj} = \sqrt{ 2 E^{\rm jet1}_{T} E^{\rm jet2}_{T} \left[
                    \cosh(\eta^{\rm jet1}-\eta^{\rm jet2})
                   -\cos(\phi^{\rm jet2}-\phi^{\rm jet2}) \right], } \]

where $\phi_1^{\rm jet}$ and $\phi_2^{\rm jet}$ are the azimuthal angles of the two jets. 
For jets back-to-back in azimuthal angle and of equal transverse energy, $E_T^{\rm jet}$, 
the dijet mass is related to the scattering angle by

\begin{equation}
M^{\rm jj} = \frac{2E^{\rm jet}_T}{\sqrt{1-|\cts|^2}}.
\label{eq:mjjapprox}
\end{equation}

When the minimum $E_T^{\rm jet}$ is taken to be 12.5~GeV, the 
average of the minimum transverse energies of the two jets, the requirement on the minimum 
dijet mass up to a given value of the scattering angle can be deduced from Eq.~(\ref{eq:mjjapprox}). 
Simulation studies show that 
the choice of cut dictated by this approximation does indeed eliminate any bias from the choice 
of transverse-energy cuts~\cite{heaphy:phd:2001}. A further cut on the boost of the dijet system 
in the laboratory frame, $\bar{\eta} = (\eta^{\rm jet1}+\eta^{\rm jet2})/2$, of 
$0.1<\bar{\eta}<1.3$ was also applied. This ensures that the phase space is uniform as a function 
of $|\cts|$, so that any shape seen in the measured distributions is attributable to the dynamics 
and not biased by the cuts imposed.

\section{Monte Carlo models}
\label{sect:mc}

The acceptance and the effects of detector response were determined using samples of simulated 
events. The programs {\sc Herwig~6.1}~\cite{cpc:67:465} and {\sc Pythia~6.1}~\cite{cpc:82:74}, 
which implement the leading-order matrix elements, followed by parton showers and hadronisation,
were used. The {\sc Herwig} and {\sc Pythia} generators differ in the details of the implementation 
of the leading-logarithmic parton-shower models. They also use different hadronisation models:  
{\sc Herwig} uses the cluster~\cite{np:b238:492} model and {\sc Pythia} uses the Lund 
string~\cite{prep:97:31} model. Direct and resolved events were generated 
separately. For all generated events, the ZEUS detector response was simulated in detail using a 
program based on {\sc Geant}~3.13~\cite{tech:cern-dd-ee-84-1}.

Parameters tuned to HERA data~\cite{proc:hera:1995:613} were used 
in the generation of the {\sc Herwig} sample. The GRV-LO~\cite{pr:d45:3986,*pr:d46:1973} and 
CTEQ4L~\cite{pr:d55:1280} set of PDFs were used for the 
photon and proton, respectively. For the {\sc Pythia} generator, the parameters were chosen to be 
consistent with fits to jet data from both HERA and LEP~\cite{proc:photon:1999:taylor}. Here, 
the SaS-2D~\cite{pl:b376:193} and GRV94-LO~\cite{zfp:c67:433} set of PDFs 
were used for the photon and proton, respectively. Multiparton interactions (MPI) were also included 
with a minimum transverse momentum of the secondary scatter of 2.0~GeV~\cite{pr:d36:2019}. However,
at the high transverse energies studied here, the effects of the ``underlying event'' are 
small: models with or without MPI describe the data equally well.

\section{NLO QCD calculations}
\label{sec:calc}

Many calculations of jet photoproduction at NLO 
exist~\cite{np:b467:399,*np:b507:295,np:b507:315,pr:d56:4007,pr:d57:5555,zfp:c76:67,epjdirect:c1:1,epj:c17:413}, 
all of which have been compared with each other and agree to within 
$(5-10)\%$~\cite{epj:c17:413,proc:mc:1998:171}. The calculation used here is that of Frixione and 
Ridolfi~\cite{np:b467:399,*np:b507:295,np:b507:315}, which employs the subtraction 
method~\cite{np:b178:421} for dealing with the collinear and infra-red divergences. The number of 
flavours was set to 5 and the renormalisation and factorisation scales, $\mu$, were set to half 
the sum of the transverse energies of the final-state partons, $E_T/2$. Two different 
parameterisations of the photon parton density were used\footnote{The GS96-HO~\cite{np:b489:405} 
parameterisation was not used because it has been shown that the available code does not reproduce 
the published results~\cite{proc:photon:1997:vogt}.}: 
GRV-HO~\cite{pr:d45:3986,*pr:d46:1973} and AFG-HO~\cite{zfp:c64:621}. The parton densities in 
the proton were parameterised using CTEQ5M1~\cite{pr:d55:1280}; the value \mbox{$\alpha_s (M_Z) = 0.118$} 
used therein was adopted for the central prediction. The parameterisation MRST99 
\mbox{($\alpha_s (M_Z) = 0.1175$)} was also considered. Parameter settings in the NLO calculation were 
varied to test the stability of the theoretical predictions, as discussed in 
Section~\ref{sec-uncert}. 

The NLO QCD predictions were corrected for hadronisation effects using a bin-by-bin procedure 
according to $d\sigma = d\sigma^{\rm NLO} \cdot C_{\rm had}^{-1}$, where 
$d\sigma^{\rm NLO}$ is the cross section for partons in the final state of the NLO calculation. 
The hadronisation correction factor, \mbox{$C_{\rm had}$ $\equiv (1+\delta_{{\rm had}})^{-1}$}, was 
defined as the ratio of the dijet cross sections before and after the hadronisation process, 
$C_{\rm had}= d\sigma_{\rm MC}^{\rm partons}/d\sigma_{\rm MC}^{\rm hadrons}$. The value of 
$C_{\rm had}$ was taken as the mean of the ratios obtained using the {\sc Herwig} and {\sc Pythia} 
predictions. The hadronisation correction, $\delta_{\rm had}$, was generally below $10\%$ in each 
bin except at 
the edges of phase space (see Tables~\ref{table-cos}--\ref{table-xgamma}). There are significant 
migrations in \ \xgo \ for \xgo \ $>$ 0.7; however, the migrations to lower  values are small.

\section{Energy corrections}
\label{energycorr}

Kinematic variables and jets were reconstructed using a combination of track and calorimeter 
information that optimises the resolution of reconstructed kinematic variables~\cite{briskin:phd:1998}. 
The selected tracks and calorimeter clusters are referred to as Energy Flow Objects (EFOs).  

The addition of track information to the CAL information reduces the sensitivity to energy 
losses in inactive material in front of the CAL, and exploits the good momentum and angular 
resolution of the tracking for low-momentum particles. The energies of particles for which no track 
information was available (e.g. neutral particles), or for which the calorimeter energy resolution 
was better than that of the tracking (e.g. at the highest energies), were measured using CAL 
information. These energies were corrected for losses in the inactive material as discussed in a 
previous publication~\cite{epj:c11:35}. Conservation of energy and momentum in neutral current 
(NC) DIS events was exploited to determine the required energy corrections~\cite{ochs:phd:2001} 
by balancing the scattered positron with the hadronic final state. This was performed independently 
for data and simulated event samples. The EFOs thus corrected were used both to reconstruct jets 
and to determine kinematic variables. Comparisons of kinematic variables for data and simulated 
events led to the assignment of a 1\% correlated systematic uncertainty in the transverse jet 
energies and in the hadronic variables~\cite{ochs:phd:2001}. The improved precision in this 
uncertainty compared with the previous measurement~\cite{epj:c11:35} was obtained from an 
increased data sample, better selection requirements and improved parameterisations of the energy 
losses. In the overlapping kinematic region, the total cross section measured here is $6\%$ lower 
than the previous measurement~\cite{epj:c11:35}, but within the quoted uncertainties arising from 
the uncertainty in the jet energy scale.

\section{Event selection}
\label{selection}

After applying the energy corrections described in Section~\ref{energycorr}, dijet events were 
selected offline by using the following procedures and cuts designed to remove sources 
of background:

\begin{itemize}

\item the $k_T$ clustering algorithm was applied to the corrected EFOs. Events were selected 
      in which at least two jets were found with $-1<\eta^{\rm jet1,2}<2.4$, 
      $E_{T}^{\rm jet1} >$~14 GeV and $E_{T}^{\rm jet2} >$~11 GeV; 

\item to remove background due to proton beam-gas interactions and cosmic-ray showers, the 
      longitudinal position of the reconstructed vertex was required to be in the range 
      $|Z_{\rm vertex}|<40$~cm;

\item to remove background due to charged current DIS events and cosmic-ray showers, a cut on the 
      relative transverse momentum of $p_T/\sqrt{E_T} < 1.5\sqrt{\rm GeV}$ was made, 
      where $p_T$ and $E_T$ are, respectively, 
      the measured transverse momentum and transverse energy of the event;

\item NC DIS events with a scattered positron candidate in the CAL were removed by 
      cutting~\cite{pl:b322:287} on the inelasticity, $y$, which is estimated from 
      \mbox{$y_e=1-\frac{E_e^\prime}{2E_e}(1-\cos\theta_e^\prime)$}, where 
      $E_e^\prime$ and $\theta_e^\prime$ are the energy and polar angle, respectively, of the 
      scattered positron candidate. Events were rejected if $y_e < 0.85$;

\item the requirement $0.2<y_{\rm JB}<0.85$ was imposed, where $y_{\rm JB}$ is the 
      estimator of $y$ measured from the CAL energy deposits according to the Jacquet-Blondel 
      method~\cite{proc:epfacility:1979:391}. The upper cut removed NC DIS events where 
      the lepton was not identified and which therefore have a value of $y_{\rm JB}$ close to 1.
      The lower cut removed proton beam-gas events which have a low value of $y_{\rm JB}$. These 
      requirements on $y_{\rm JB}$ restrict the photon-proton centre-of-mass energy to be in the 
      range 134 to 277~GeV.

\end{itemize}

The cuts on $y_e$ and $y_{\rm JB}$ reduced the background from DIS events to less than $0.5\%$ and 
restricted the range of the virtuality of the exchanged photon to $Q^2 < 1~{\rm GeV^2}$, with a median 
value of about 10$^{-3}$~${\rm GeV^2}$. After these requirements, 62573 events with two or more jets 
remained in the sample; 2919 of these events had a third jet of transverse energy greater than 
11~GeV in the range $-1<\eta^{\rm jet3}<2.4$. For the measurements of the cross section as a 
function of $\cts$, the following additional requirements were imposed:

\begin{itemize}

\item the dijet scattering angle was restricted to be in the range $|\cts|<0.8$;

\item the invariant mass of the two jets of highest transverse energy was required to 
      satisfy $M^{\rm jj} > $ 42~GeV;

\item the boost of the dijet system was required to be in the range $0.1<\bar{\eta}<1.3$.

\end{itemize}

These cuts reduced the sample to 10811 events.

\section{Event characteristics}

For the high transverse energies studied here, it has been shown 
previously~\cite{epj:c1:109,epj:c11:35} that the transverse energy flow around jets is generally 
well described by the simulation. This agreement is maintained in the more-forward region 
studied here, with {\sc Pythia} giving a description similar to that of {\sc Herwig}.

\begin{figure}[htb]
\begin{center}
~\epsfig{file=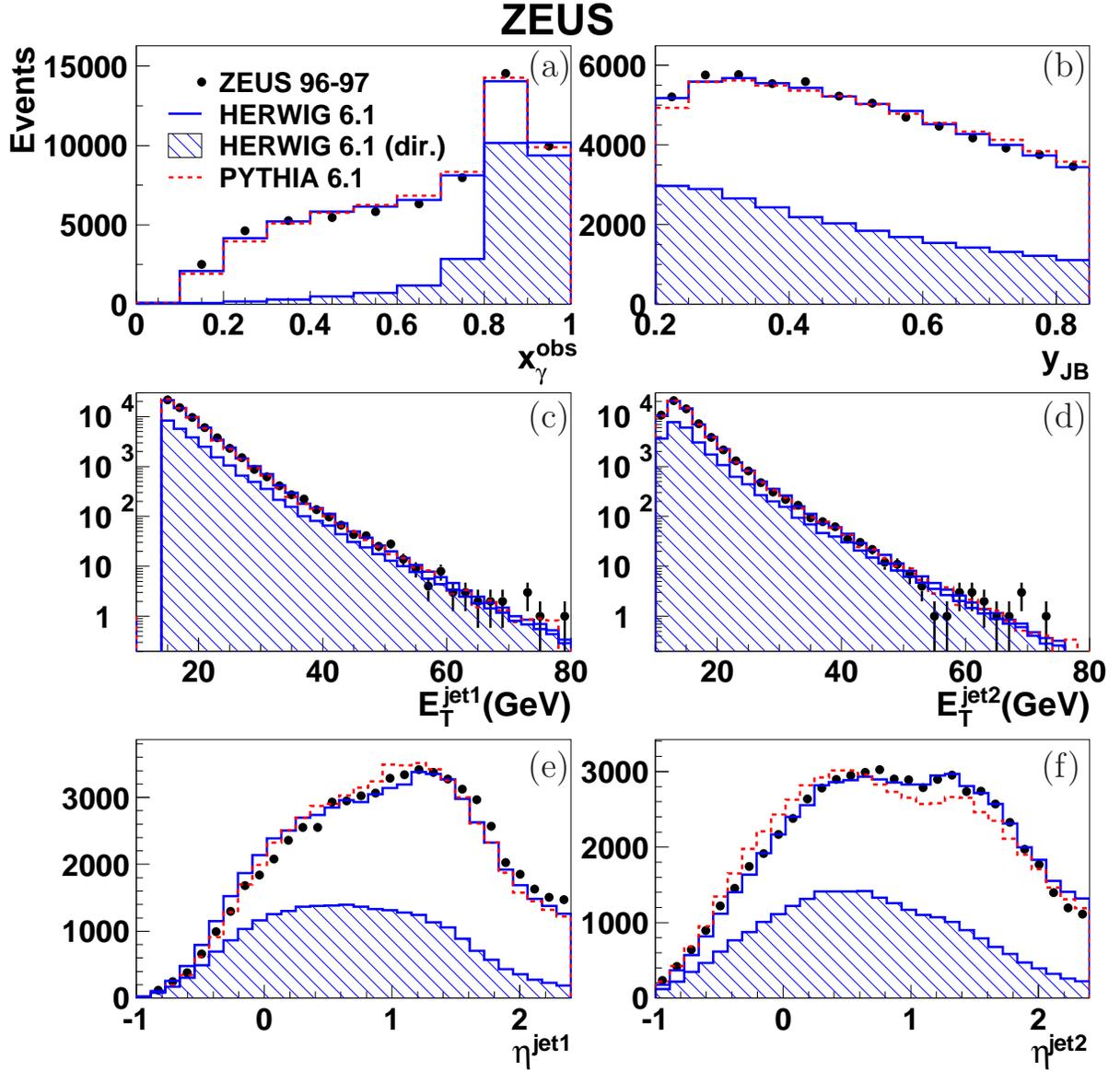,height=16cm}
\put(-240,426){\makebox(0,0)[tl]{\large (a)}}  
\put(-29,426){\makebox(0,0)[tl]{\large (b)}}  
\put(-240,282){\makebox(0,0)[tl]{\large (c)}}  
\put(-29,282){\makebox(0,0)[tl]{\large (d)}}  
\put(-240,138){\makebox(0,0)[tl]{\large (e)}}  
\put(-29,138){\makebox(0,0)[tl]{\large (f)}}  
\end{center}
\caption{Comparison of data and MC simulation for (a) \xgo, (b) $y_{\rm JB}$, 
(c) $E_{T}^{\rm jet1}$, 
(d) $E_{T}^{\rm jet2}$, (e) $\eta^{\rm jet1}$ and (f) $\eta^{\rm jet2}$. The data are shown as 
points compared to {\sc Herwig} (solid line) and {\sc Pythia} (dashed line). Also shown 
is the LO component of direct photon processes in {\sc Herwig} (hatched area). The simulated sample
is normalised to the data and the fraction of direct and resolved photon processes combined 
according to a $\chi^2$-fit to the \xgo \ distribution in (a).}
\label{fig:mc}
\end{figure}

Kinematic distributions in the data are compared to the two simulation programs in 
Fig.~\ref{fig:mc}. The simulated distributions were fit to the data distribution in \xgo, shown 
in Fig.~\ref{fig:mc}(a), by varying the fractions of direct and resolved processes and minimising 
the $\chi^2$. The simulations generally describe the data well for all variables, although 
some discrepancies are seen. 
The shape of the distribution in $y_{\rm JB}$ is better described by the {\sc Herwig} prediction, 
although the description by the {\sc Pythia} simulation is adequate. Both simulations have similar 
distributions for the transverse energies of the jets and describe the data well. However, neither 
gives a good description of the pseudorapidity of the jet of highest transverse energy, whereas the 
pseudorapidity distribution of the second jet is well described by {\sc Herwig}. Figure~\ref{fig:mc} 
also shows that according to the {\sc Herwig} simulation, the proportion of resolved photon events 
decreases with increasing \xgo \ and $E_T^{\rm jet}$ and increases with increasing $y_{\rm JB}$ and 
$\eta^{\rm jet}$.

Since {\sc Herwig} gives a better overall description of the data than {\sc Pythia}, 
it was chosen as the primary MC generator to correct the data. The correction was performed using 
the bin-by-bin 
method, in which the correction factor, as a function of an observable ${\mathcal O}$ in a 
given bin $i$, is $C_i({\mathcal O})=N_i^{\rm had}({\mathcal O})/N_i^{\rm det}({\mathcal O})$. 
The variable $N_i^{\rm had}({\mathcal O})$ is the number of events passing the kinematic 
requirements on the hadronic final state described in Section~\ref{photon_structure} and 
$N_i^{\rm det}({\mathcal O})$ is the number of reconstructed events passing the selection 
requirements as detailed in Section~\ref{selection}. Both numbers were computed using the MC
generators described in Section~\ref{sect:mc}. For distributions as a function of \xgo, the correction 
factors lie in the range 0.85$-$1.25.

\section{Experimental and theoretical uncertainties}
\label{sec-uncert}

\subsection{Experimental uncertainties}

The results of a detailed analysis of the possible sources of systematic uncertainty 
are listed below. Typical values for the systematic uncertainty are quoted for the 
cross sections as a function of \xgo:

\begin{itemize}

\item varying the measured jet energies by $\pm1\%$ in only the simulated sample, in 
      accordance with the uncertainty in the jet energy scale, gave an uncertainty of $\mp 5\%$, 
      increasing with $E_{T}^{\rm jet1}$;

\item correcting the data with {\sc Pythia} instead of the {\sc Herwig} generator gave an 
      uncertainty within  $\pm 9\%$ and typically $\pm 4\%$;

\item changing the cuts on $E_{T}^{\rm jet1}$ and $E_{T}^{\rm jet2}$ in both the data and 
      simulated samples by the value of the average resolution ($\sim 9\%$) gave an uncertainty of 
      $\pm5\%$;

\item changing the cuts on $y_{\rm JB}$ in both the data and simulated samples by the value of the 
      resolution ($\sim 0.03$ at low $y_{\rm JB}$ and $\sim 0.05$ at high $y_{\rm JB}$) 
      gave an uncertainty of less than $\pm2\%$;

\item changing the cuts on $\eta^{\rm jet1,2}$ in both the data and simulated 
      samples by the value of the resolution ($\sim 0.04$) gave an uncertainty of $\pm0.5\%$;

\item varying the cuts to remove DIS and beam-gas backgrounds in both data 
      and simulated samples gave a total uncertainty of less than $\pm1\%$.

\end{itemize}

In addition to the above, the cuts made to evaluate the cross section as a function of $|\cts|$ 
also lead to sources of systematic uncertainty. The following were evaluated, with typical 
uncertainties quoted:

\begin{itemize}

\item changing the cuts on $M^{\rm jj}$ in both the data and simulated samples by the 
      value of the average resolution ($\sim 8\%$) gave an uncertainty of $\pm5\%$;

\item changing the cuts on $\bar{\eta}$ in both the data and simulated samples by the value of the 
      resolution ($\sim 0.04$) gave an uncertainty of $\pm2\%$;

\item changing the cuts on \xgo \ in both the data and simulated samples by the value of the 
      resolution ($\sim 0.04$) gave an uncertainty of $\pm4\%$. This systematic uncertainty  
      also contributes to the measurements of the cross-sections $d\sigma/dE_{T}^{\rm jet1}$ 
      and $d\sigma/d\eta^{\rm jet2}$.

\end{itemize}

The uncertainty in the cross sections due to the jet energy scale uncertainty is correlated 
between bins and is therefore displayed separately as a shaded band in the figures. All other 
systematic uncertainties were added in quadrature. In addition, an overall normalisation 
uncertainty of $1.6\%$ from the luminosity determination is not included in either the figures or 
tables.

As a cross check, cross sections were obtained with an iterative matrix-unfolding 
technique\cite{nim:a362:487}, using Bayes' theorem. The resultant cross sections were found to be 
consistent with those using the standard bin-by-bin procedure~\cite{heaphy:phd:2001}.

\subsection{Theoretical uncertainties}
\label{sec:theo_uncert}

The NLO QCD predictions for the dijet cross section are affected by the systematic uncertainties 
listed below. Typical values for the systematic uncertainty are quoted for the 
cross sections as a function of \xgo:

\begin{itemize}

\item the uncertainty due to terms beyond NLO, estimated by varying 
      $\mu$ between $E_T$ and $E_T/4$, is $\pm (10-20)\%$, increasing with 
      decreasing \xgo;

\item the uncertainty due to the hadronisation correction, estimated as half the 
      spread between the $\delta_{\rm had}$ values obtained using the 
      {\sc Herwig} and {\sc Pythia} models, is $\pm (2-3)\%$;

\item the uncertainty due to the value of $\alpha_s(M_Z)$, estimated by repeating the 
      calculations using the CTEQ4 series of PDFs determined using  
      values of $\alpha_s(M_Z) =$ 0.113, 0.116 and 0.119, is $\pm (5-8)\%$.

\end{itemize}

The above systematic uncertainties are largely independent of the choice of photon PDF and 
were added in quadrature to give the total uncertainty on the predictions in each case. 
Differences between parameterisations of the photon and proton parton 
densities are discussed in the comparison to the measured data in Section~\ref{sec-res}.

\section{Results}
\label{sec-res}

\subsection{Probing the matrix elements}

\begin{figure}[htb]
\begin{center}
~\epsfig{file=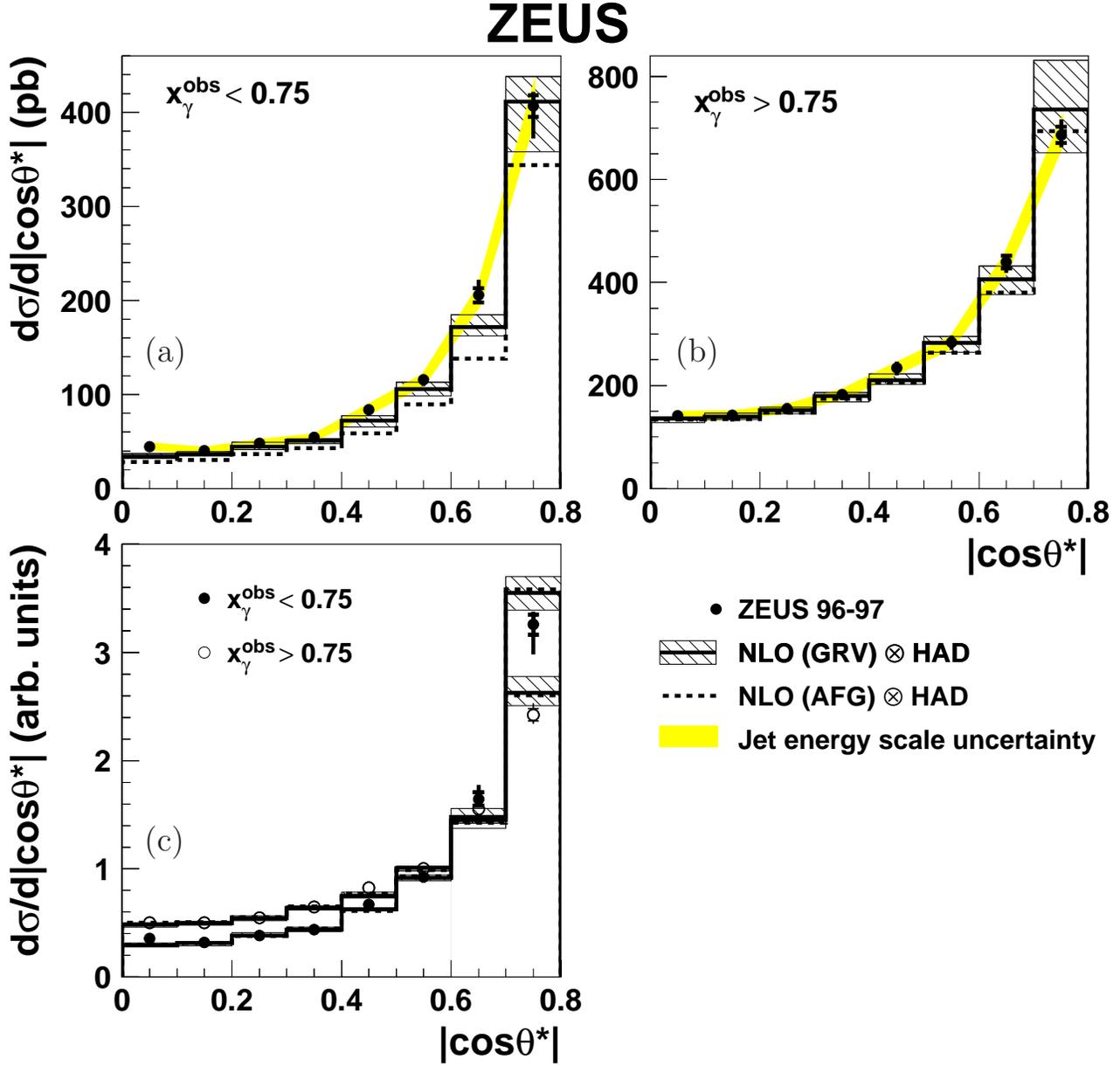,height=17cm}
\put(-421,326){\makebox(0,0)[tl]{\large (a)}}  
\put(-191,326){\makebox(0,0)[tl]{\large (b)}} 
\put(-421,114){\makebox(0,0)[tl]{\large (c)}}  
\caption{
Measured cross sections as a function of $|\cts|$ for (a) \xgo \ $<0.75$ and (b) 
\xgo \ $>0.75$, compared to NLO predictions. The data are shown with statistical errors (inner 
bars) and statistical and systematic uncertainties added in quadrature (outer bars). 
The uncertainty due to that of the jet energy scale is shown as the shaded band. 
The NLO prediction corrected for hadronisation effects is shown calculated using the GRV-HO 
and CTEQ5M1 PDFs for the photon and proton, respectively, and the scale set to $E_T/2$ 
(solid line). 
The hatched band represents the quadratic sum of the theoretical uncertainties as discussed in 
Section~\ref{sec:theo_uncert}. The prediction using the AFG-HO photon PDF is 
also shown (dashed line). In (c) the cross sections are area-normalised and the data shown for 
$\xgo \ <0.75$ (solid points) and $\xgo \ >0.75$ (open circles).}
\label{fig:cos}
\end{center}
\end{figure}

The dijet cross section as a function of $|\cts|$, is given in 
Table~\ref{table-cos} and shown in Fig.~\ref{fig:cos}. The data are shown separately for 
\xgo \ $<$ 0.75 (Fig.~\ref{fig:cos}(a)) and \xgo \ $>$ 0.75 (Fig.~\ref{fig:cos}(b)) and 
compared to NLO predictions. For \xgo \ $<$ 0.75, the measured cross section lies above the 
NLO prediction using GRV-HO for the photon PDF by an average of $(10-15)\%$. Considering the 
theoretical and experimental uncertainties, both of $(5-10)\%$, the NLO prediction gives a 
reasonable description of the data. The predictions using the AFG-HO parameterisation for the 
photon give a lower cross section than that of GRV-HO, and are thus 
around $(20-25)\%$ lower than the data. For \mbox{\xgo \ $>0.75$}, the NLO prediction is in  
agreement with the measured cross section.  

In Fig.~\ref{fig:cos}(c), the shapes of the data and NLO distributions are compared. The 
predictions give a generally good description of the data; the shapes of the predictions when 
using the GRV-HO and AFG-HO parameterisations are similar. The data at low \xgo \ rise more 
rapidly at high $|\cts|$ than those at high \xgo. This is consistent with a difference in the 
dominant propagators, as observed in a previous publication~\cite{pl:b384:401}; this is seen 
here at higher energies and masses. The agreement in shape of these distributions, which are 
sensitive to the matrix elements, demonstrates that also in this high-mass region the dynamics 
of the short-distance process is understood.

\subsection{Cross sections in $E_{T}^{\rm jet1}$ and $\eta^{\rm jet2}$}

\begin{figure}[htb]
\begin{center}
~\epsfig{file=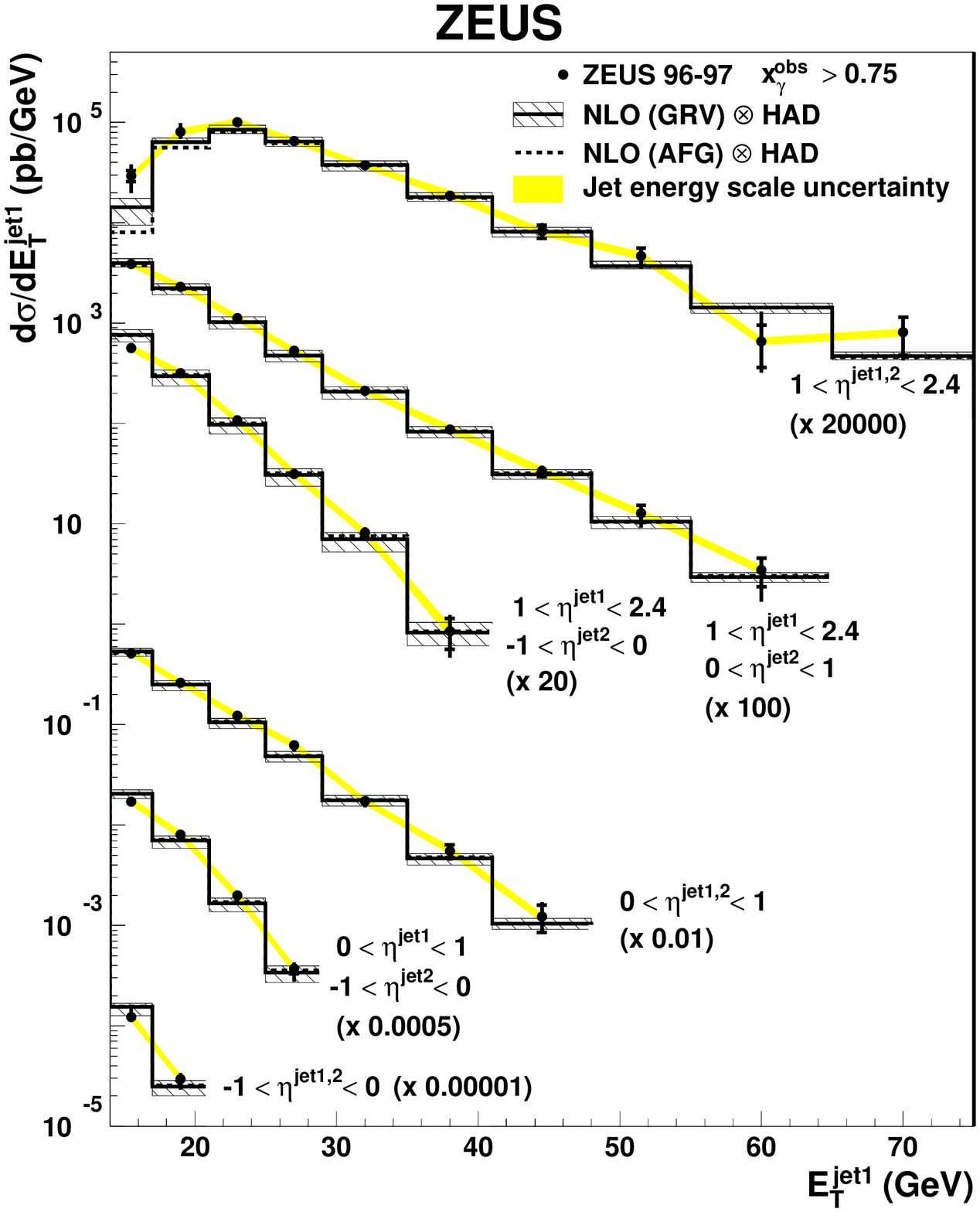,height=20cm}
\caption{Measured cross section as a function of $E_{T}^{\rm jet1}$ for events with $\xgo \ > 0.75$. 
The measurement is divided into six regions of the pseudorapidities of the jets. 
The cross sections are multiplied by the scale factor indicated in brackets so that all 
regions can be displayed in the same figure. For further details, see the caption to 
Fig.~\ref{fig:cos}.}
\label{fig:et_xsec_highxg}
\end{center}
\end{figure}

\begin{figure}[htb]
\begin{center}
~\epsfig{file=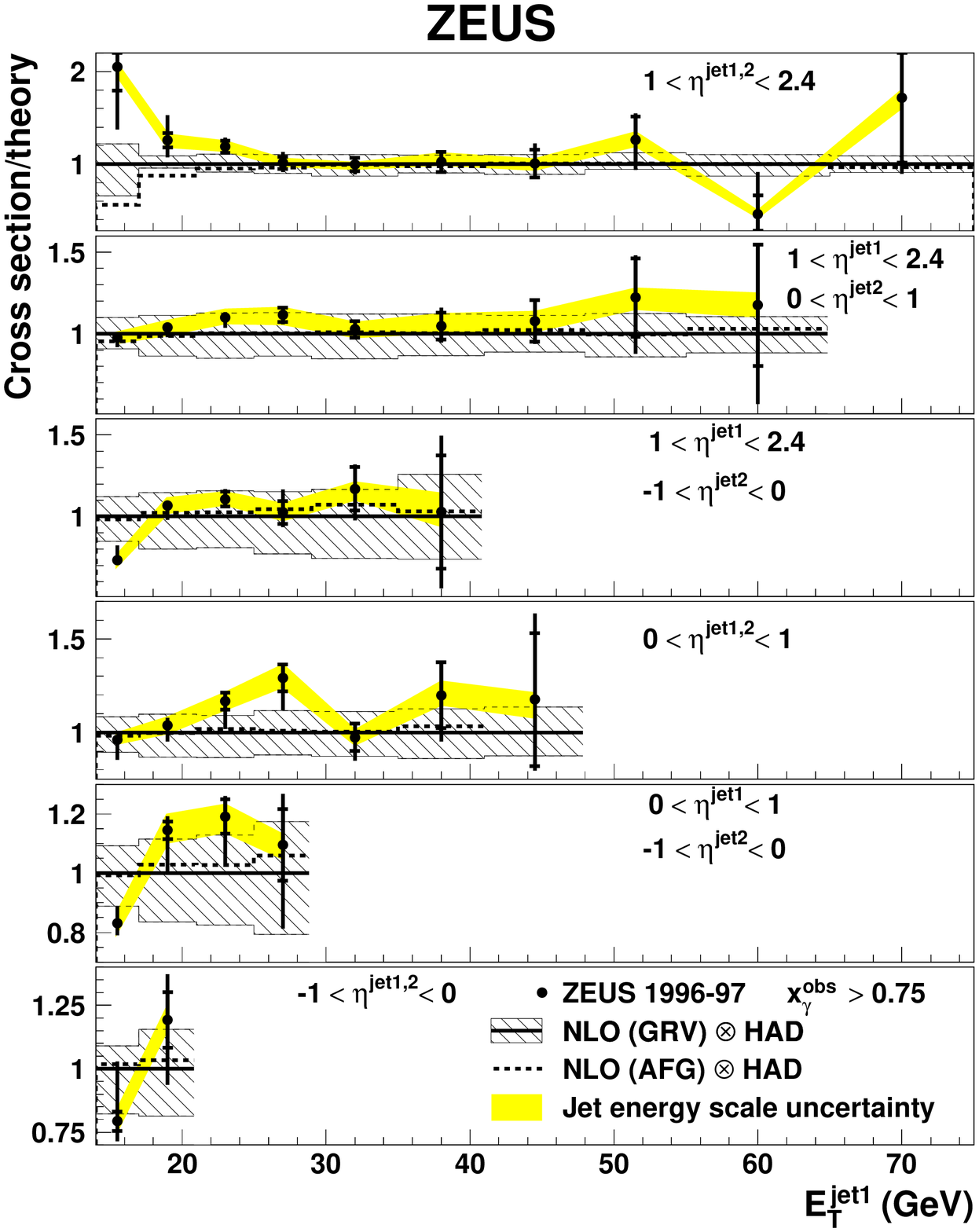,height=20cm}
\caption{Ratio of cross sections to the central theoretical prediction as a function of 
$E_{T}^{\rm jet1}$ for events with $\xgo \ > 0.75$. The measurement is divided into six 
regions of the pseudorapidities of the jets. For further details, see the caption to 
Fig.~\ref{fig:cos}.}
\label{fig:et_ratio_highxg}
\end{center}
\end{figure}

\begin{figure}[htb]
\begin{center}
~\epsfig{file=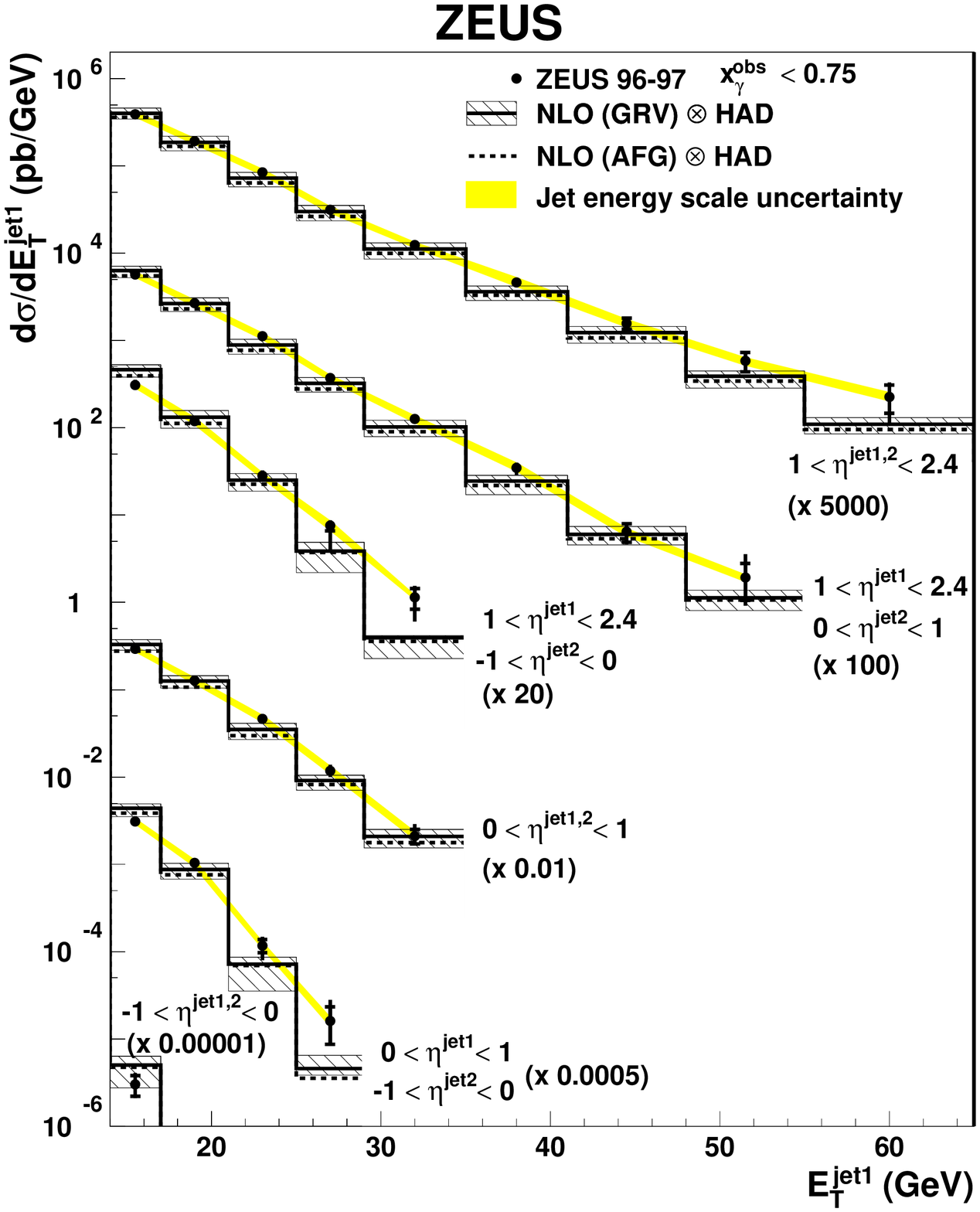,height=20cm}
\caption{Measured cross section as a function of $E_{T}^{\rm jet1}$ for events with $\xgo \ < 0.75$. 
The measurement is divided into six regions of the pseudorapidities of the jets. 
The cross sections are multiplied by the scale factor indicated in brackets so that all  
regions can be displayed in the same figures. For further details, see the caption to 
Fig.~\ref{fig:cos}.}
\label{fig:et_xsec_lowxg}
\end{center}
\end{figure}

\begin{figure}[htb]
\begin{center}
~\epsfig{file=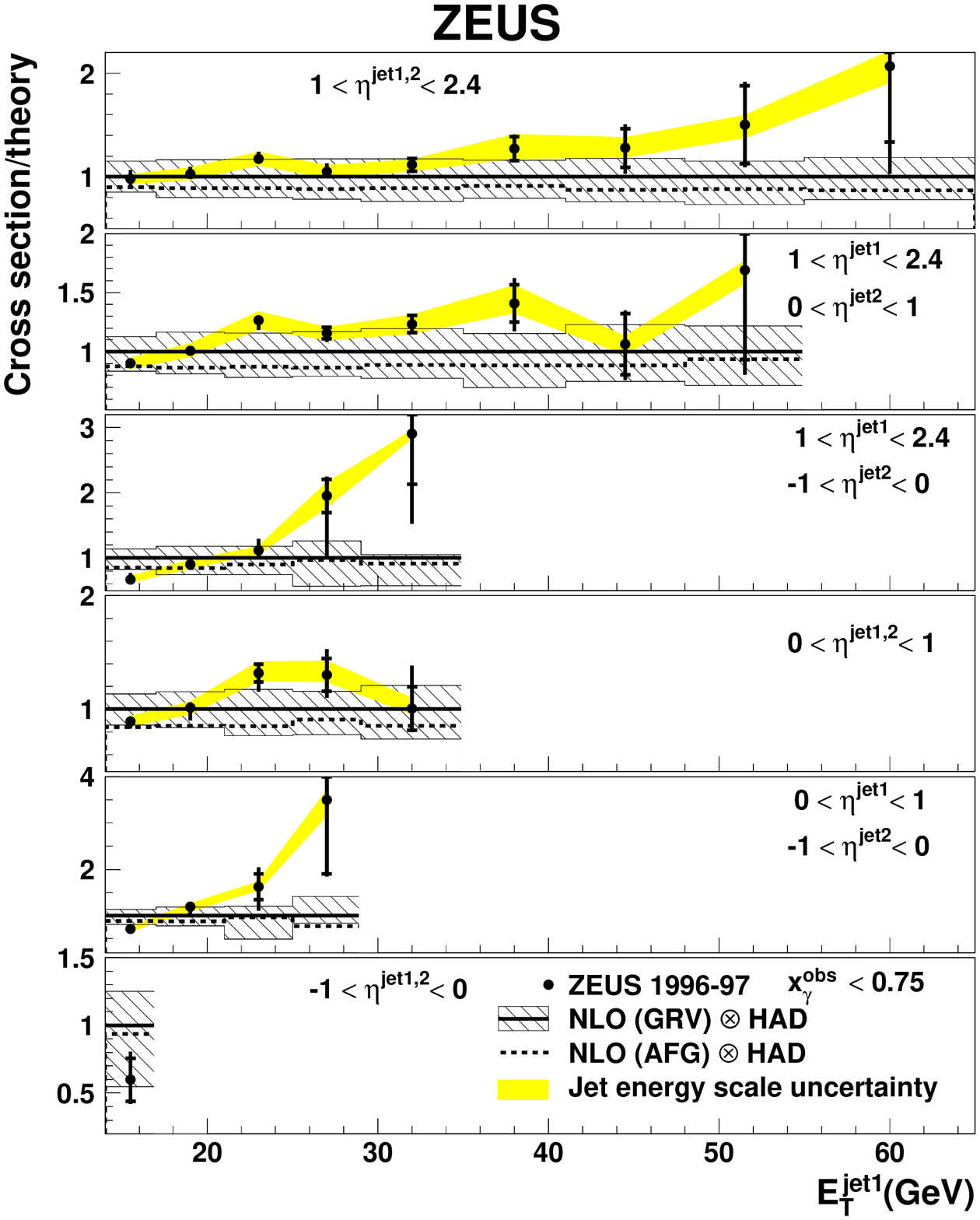,height=20cm}
\caption{Ratio of cross sections to the central theoretical prediction as a function of 
$E_{T}^{\rm jet1}$ for events with $\xgo \ < 0.75$. The measurement is divided into six 
regions of the pseudorapidities of the jets. For further details, see the caption to 
Fig.~\ref{fig:cos}.}
\label{fig:et_ratio_lowxg}
\end{center}
\end{figure}

Measurements of the dijet cross section as a function of $E_{T}^{\rm jet1}$  
are given in Tables~\ref{table-etgt} and \ref{table-etlt} and shown in 
Figs.~\ref{fig:et_xsec_highxg}--\ref{fig:et_ratio_lowxg}, in different regions of the pseudorapidities 
of the two jets for $\xgo>0.75$ and $\xgo<0.75$. The ratios of the cross sections 
to theory are shown in Figs.~\ref{fig:et_ratio_highxg} and \ref{fig:et_ratio_lowxg}. In 
Fig.~\ref{fig:et_xsec_highxg}, the measurement for $\xgo>0.75$ extends to transverse energies of 
$\sim$ 70 GeV, extending the region measured previously~\cite{epj:c11:35}. In general, the
overall description of the data by the predictions is reasonable. In particular, when the 
jets are produced in the region 
$1<\eta^{\rm jet1}<2.4$ and  $0<\eta^{\rm jet2}<1$, the cross section in 
Fig.~\ref{fig:et_xsec_highxg} falls three orders of magnitude and is well described by the NLO 
calculation. When both jets are produced in the region $1<\eta^{\rm jet1,2}<2.4$, the NLO 
prediction lies below the data at low transverse energy, although both the experimental and 
theoretical uncertainties are sizeable. In this region, the hadronisation corrections are 
significant but do not account for all the differences. This region is near the 
edge of phase space due to the cuts applied. 

At low \xgo \ (Figs.~\ref{fig:et_xsec_lowxg} and \ref{fig:et_ratio_lowxg}), the data are also 
generally well described by the NLO predictions, although a difference in shape is seen, with the 
predictions lying above the data at low transverse energy and below for $E_{T}^{\rm jet1} > 20$~GeV. 
The predictions using AFG-HO are uniformly about $15\%$ below those of GRV-HO for the entire 
range of transverse energies.

\begin{figure}[htb]
\begin{center}
~\epsfig{file=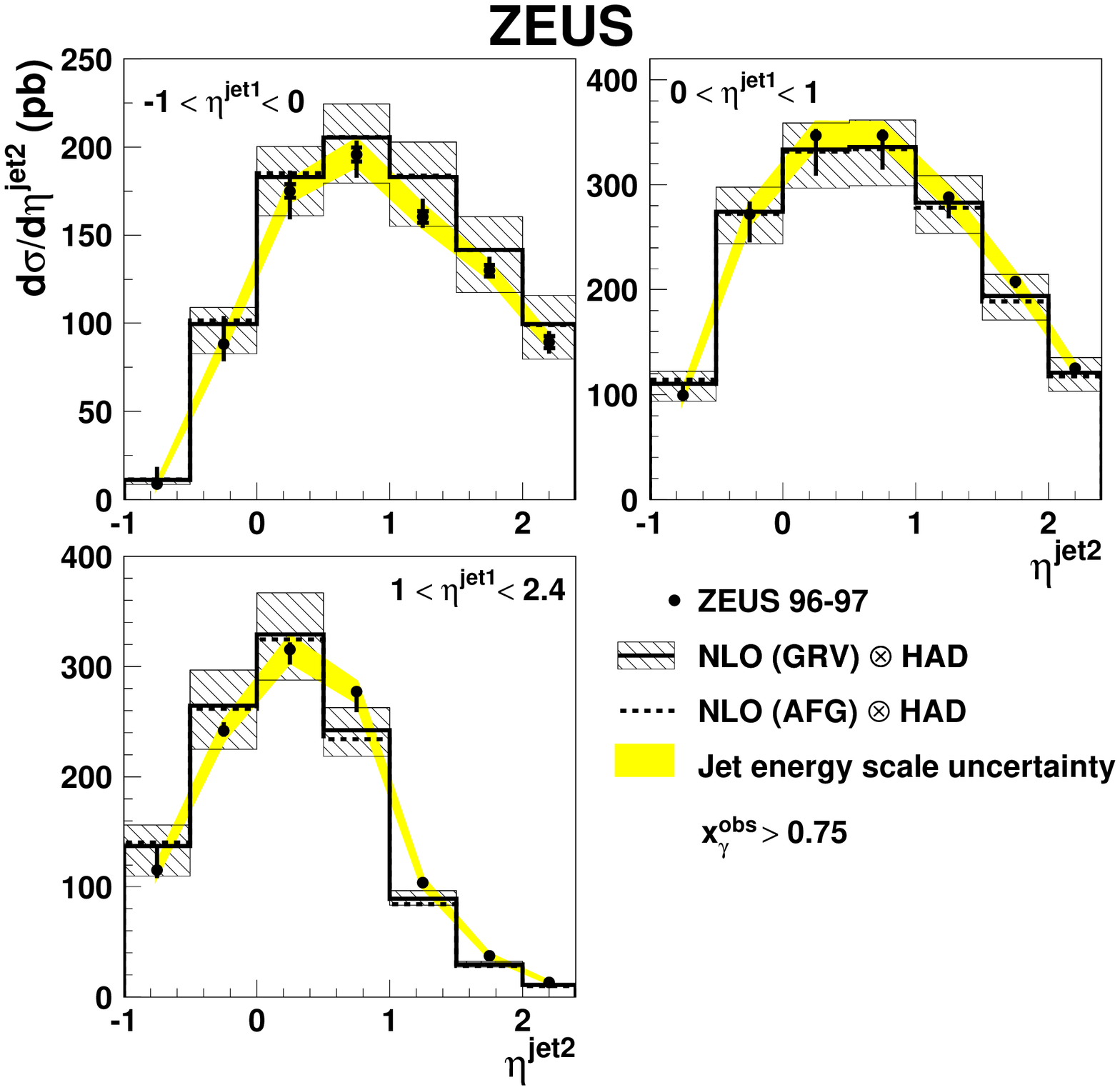,height=17cm}
\caption{Measured cross section as a function of $\eta^{\rm jet2}$ for events with $\xgo \ > 0.75$. 
The measurement is divided into three regions of the pseudorapidity of the other 
jet. For further details, see the caption to Fig.~\ref{fig:cos}.}
\label{fig:eta_xsec_highxg}
\end{center}
\end{figure}

\begin{figure}[htb]
\begin{center}
~\epsfig{file=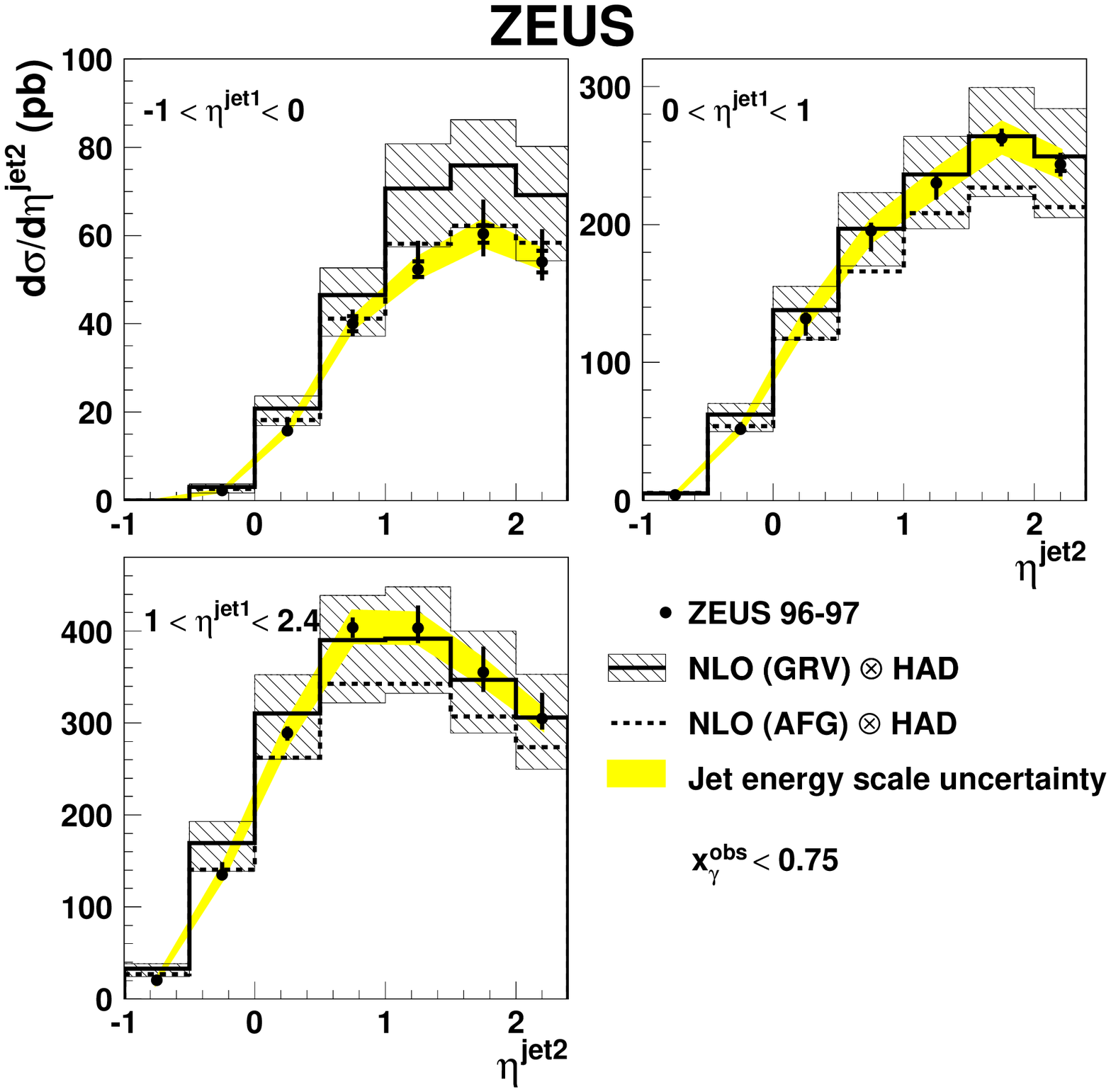,height=17cm}
\caption{Measured cross section as a function of $\eta^{\rm jet2}$ for events with $\xgo \ < 0.75$. 
The measurement is divided into three regions of the pseudorapidity of the other 
jet. For further details, see the caption to Fig.~\ref{fig:cos}.}
\label{fig:eta_xsec_lowxg}
\end{center}
\end{figure}

The pseudorapidities of the two jets are sensitive to the momentum distributions of the incoming 
partons. The cross 
section is measured as a function of the pseudorapidity of one of the jets, in different regions of 
pseudorapidity of the other, and is shown at high \xgo \ (Fig.~\ref{fig:eta_xsec_highxg}) and 
low \xgo \ (Fig.~\ref{fig:eta_xsec_lowxg}). The NLO predictions give a good description of the 
data except for $-1 < \eta^{\rm jet1} < 0$ 
for low \xgo, where the data are at or below the lower edge of the scale-uncertainty band.  
The predictions using AFG-HO lie about $(10-15)\%$ below those of GRV-HO.

\subsection{Testing the current parameterisations of the photon PDF}

\begin{figure}[htb]
\begin{center}
~\epsfig{file=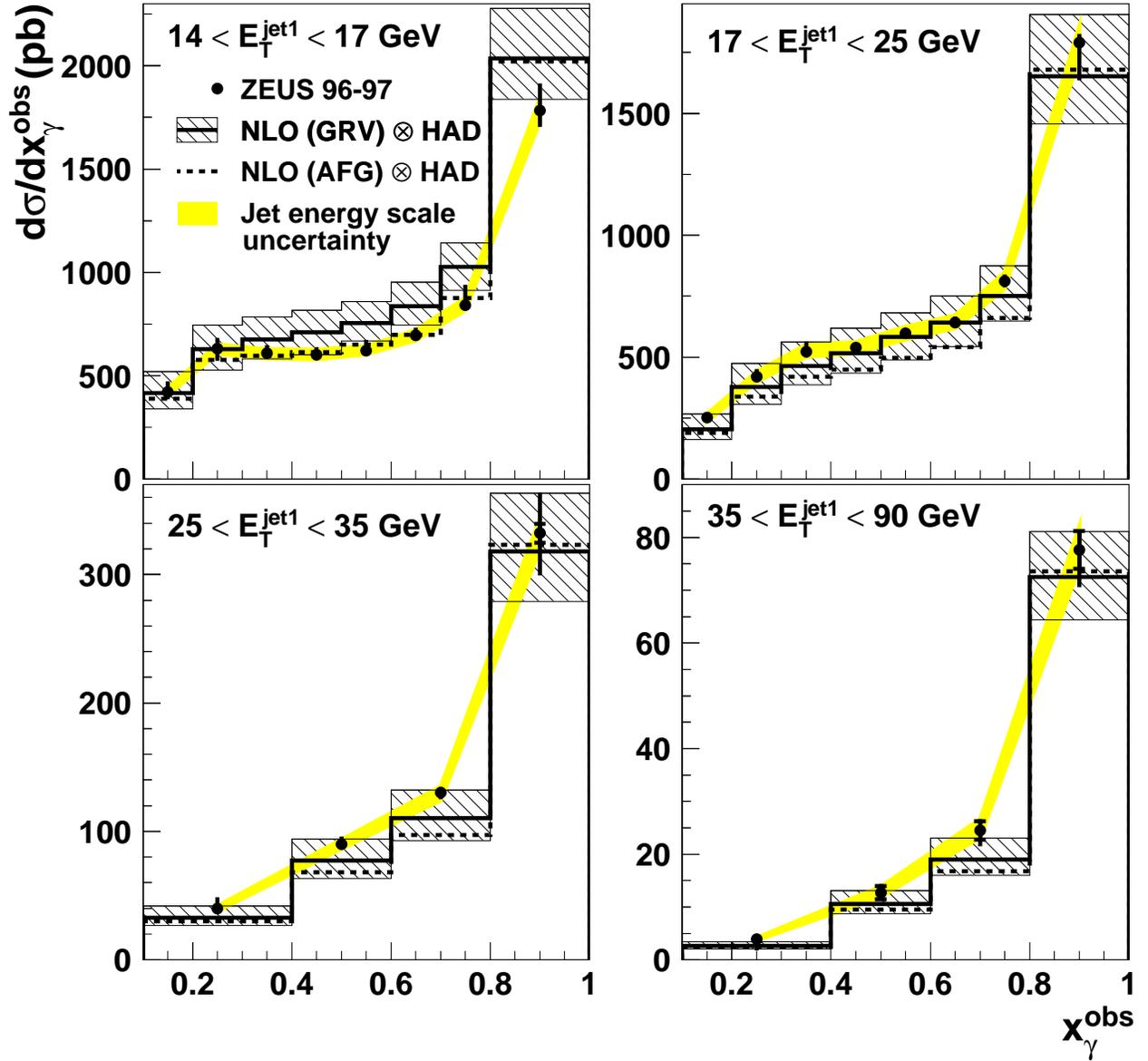,height=17cm}
\caption{Measured cross section as a function of \xgo \ in four regions of $E_{T}^{\rm jet1}$ 
compared to NLO predictions. For further details, see the caption to 
Fig.~\ref{fig:cos}.}
\label{fig:xgamma_xsec}
\end{center}
\end{figure}

\begin{figure}[htb]
\begin{center}
~\epsfig{file=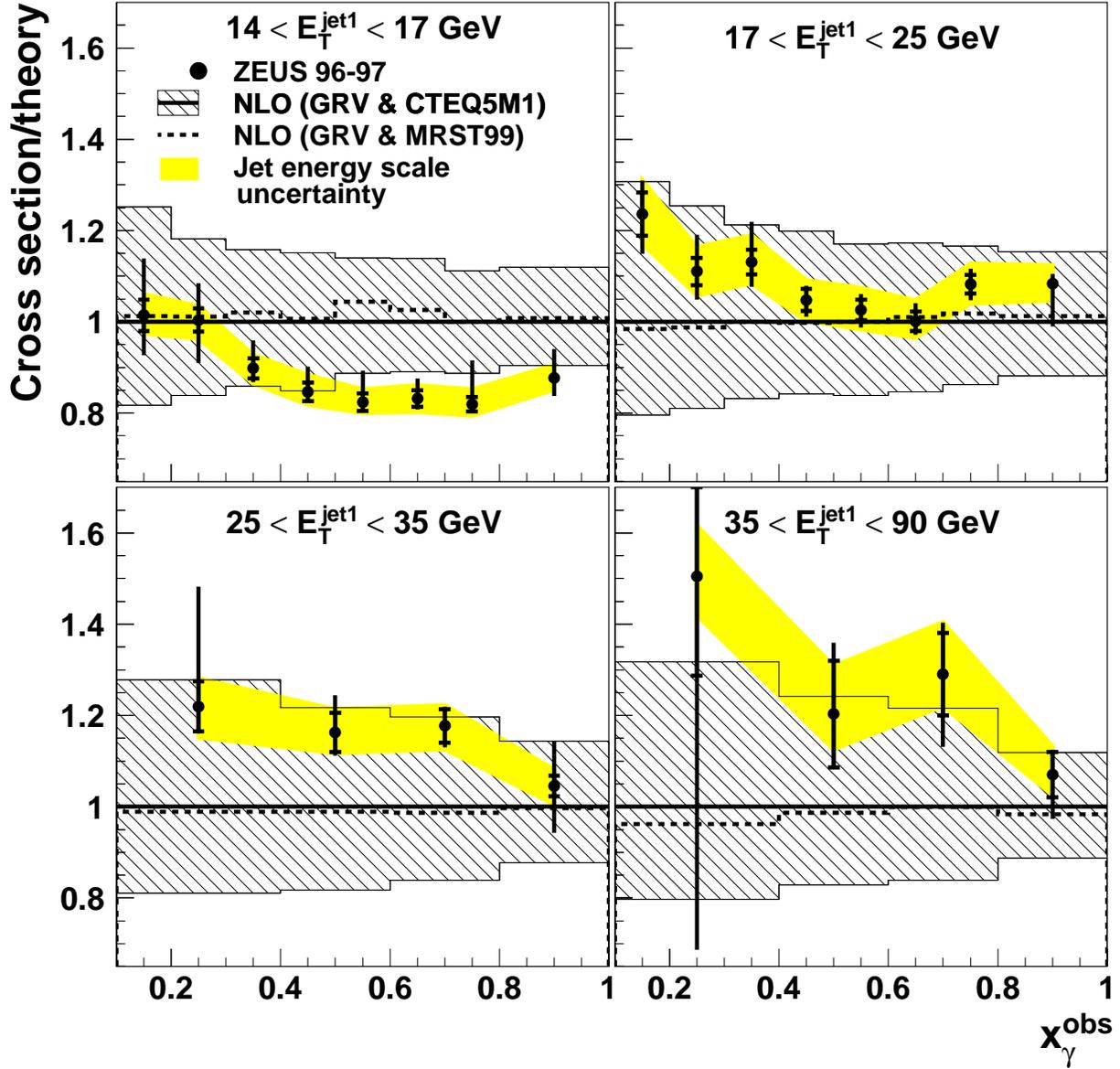,height=17cm}
\caption{Ratio of cross sections to the NLO prediction using GRV-HO and CTEQ5M1 as the photon and 
proton PDFs, respectively, and the scale set to $E_T/2$ as a function of \xgo \ in four regions of 
$E_{T}^{\rm jet1}$. The data are shown with statistical errors (inner bars) and statistical and 
systematic uncertainties added in quadrature (outer bars). The uncertainty due to that of the jet 
energy scale is shown as the shaded band. The theoretical uncertainty is shown as the hatched 
band. Predictions using MRST99 (dashed line) for the proton are also shown.}
\label{fig:xgamma_grv_ratio}
\end{center}
\end{figure}

\begin{figure}[htb]
\begin{center}
~\epsfig{file=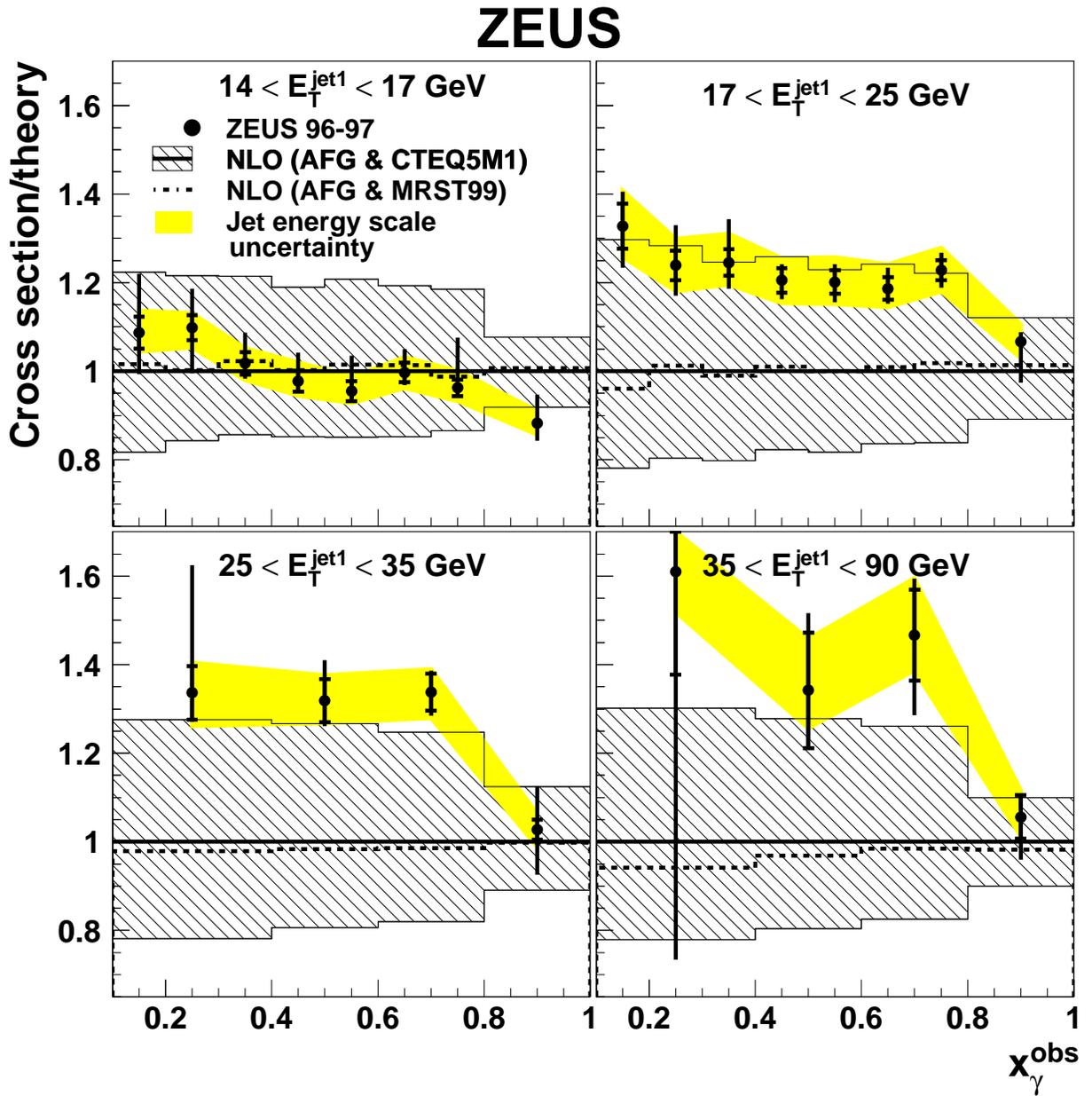,height=17cm}
\caption{Ratio of cross sections to the NLO prediction using AFG-HO and CTEQ5M1 as the photon and 
proton PDFs, respectively, and the scale set to $E_T/2$ as a function of \xgo \ in regions of 
$E_{T}^{\rm jet1}$. For further details, see the caption to Fig.~\ref{fig:xgamma_grv_ratio}.}
\label{fig:xgamma_afg_ratio}
\end{center}
\end{figure}

The cross sections and ratios of data and theory as a function of \xgo \ in regions of increasing 
transverse energy are shown in Figs.~\ref{fig:xgamma_xsec}--\ref{fig:xgamma_afg_ratio}. The 
predictions lie significantly above the data using the GRV-HO parameterisation at the lowest 
values ($14<E_{T}^{\rm jet1}<17$~GeV) of transverse energy for \xgo \ $>0.5$, but are 
increasingly below the data for values larger than 17~GeV. This trend with transverse energy is 
stronger for $\xgo<0.8$, as can be seen in Fig.~\ref{fig:xgamma_grv_ratio}. Given the 
uncertainties, the data and predictions are consistent except in the region of lowest transverse 
energy for \xgo \ $>0.5$. The dominant theoretical uncertainty, estimated from the variation of the 
renormalisation and factorisation scales, arises from the higher-order contributions not present 
in an NLO calculation. The inclusion of higher-order contributions would have to result in a 
significant change of shape of the distribution as a function of both the transverse energy and 
\xgo \ if it were to describe the data. The data in Fig.~\ref{fig:xgamma_xsec} are also compared to the 
NLO prediction using the AFG-HO photon parameterisation; the ratio of data to the theory is shown in 
Fig.~\ref{fig:xgamma_afg_ratio}. The prediction agrees with the data in the region of lowest 
transverse energy, but is below the data for the higher $E_{T}^{\rm jet1}$ bins. The predictions 
using AFG-HO are similar in shape to those using GRV-HO but are $(10-15)\%$ lower. 
Figures~\ref{fig:xgamma_grv_ratio} and \ref{fig:xgamma_afg_ratio} also show the predictions using 
the MRST99 PDFs in the proton. The differences between the predictions with CTEQ5M1 
and MRST99 are everywhere less than $5\%$. 

\subsection{Discussion}

\begin{figure}[htb]
\begin{center}
~\epsfig{file=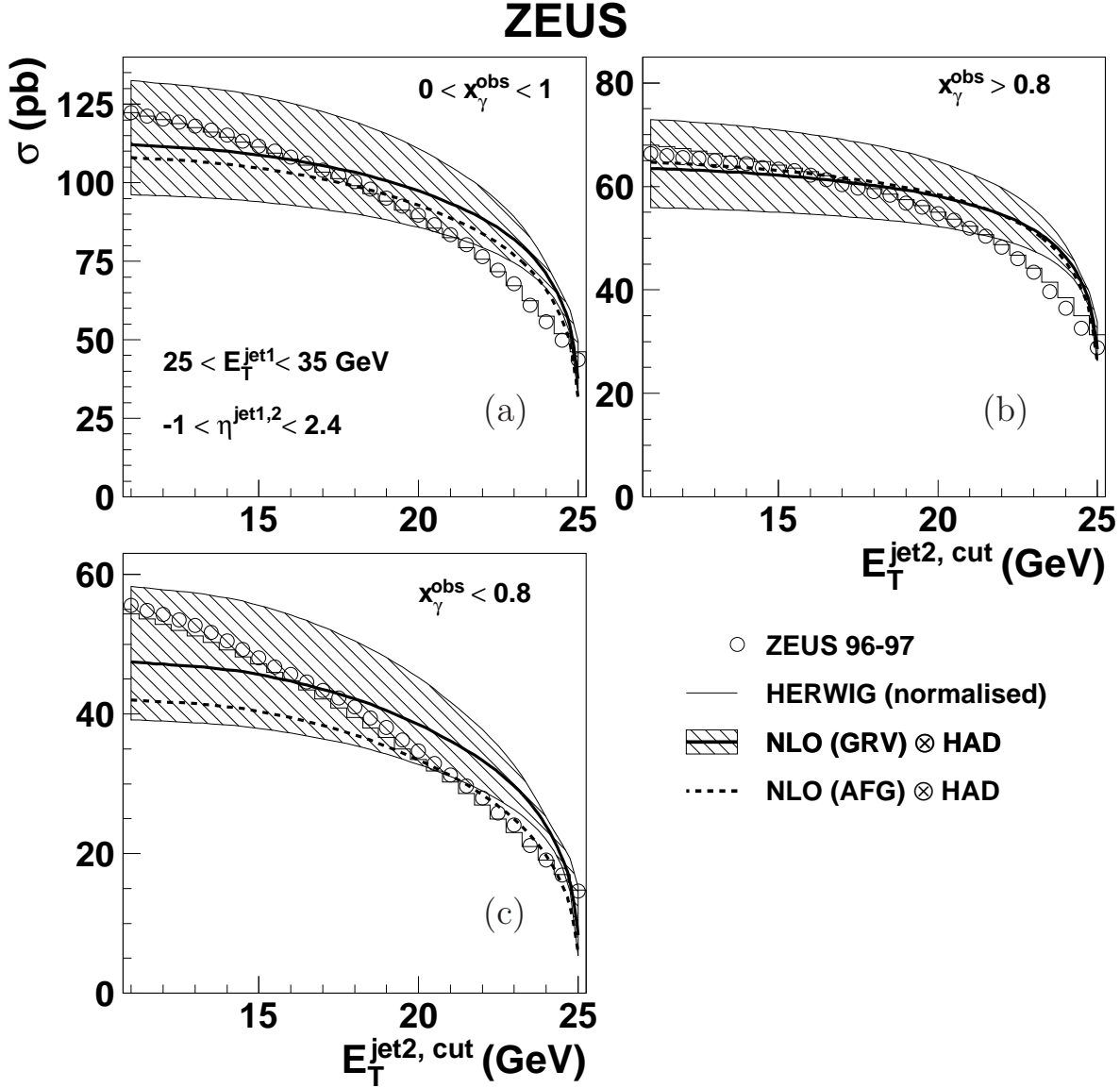,height=16cm}
\put(-261,286){\makebox(0,0)[tl]{\large (a)}}  
\put(-61,286){\makebox(0,0)[tl]{\large (b)}} 
\put(-261,84){\makebox(0,0)[tl]{\large (c)}}  
\caption{Measured cross section as a function of $E_{T}^{\rm jet2, cut}$ for a fixed range 
of transverse energy of the leading jet, $25<E_T^{\rm jet1}<35$~GeV, compared to MC 
simulation and NLO predictions for (a) 0 $< \xgo <1$, (b) $\xgo > 0.8$ and (c) $\xgo < 0.8$. 
The typical magnitude of the statistical and systematic uncertainties added in quadrature is 
$\pm 10\%$. The NLO prediction, corrected for hadronisation effects, calculated using the 
GRV-HO and CTEQ5M1 PDFs for the photon and proton, respectively, is shown as the thick solid 
line. The shaded band represents the quadratic sum of the systematic uncertainties, as 
discussed in Section~\ref{sec:theo_uncert}. The prediction using the AFG-HO photon PDF is 
shown as the dashed line. The prediction of {\sc Herwig}, calculated using the GRV-LO and 
CTEQ4L PDFs for the photon and proton, respectively, is normalised to the first point 
in the range \mbox{$0<\xgo<1$}; it is shown as the histogram.}
\label{fig:et2-cut}
\end{center}
\end{figure}

To improve the understanding of the features of the cross section in different regions of 
transverse energy, the sensitivity of the above comparisons to the value of the cut on the 
second jet has been studied. Starting at a minimum of 11~GeV, the cut on the second jet was 
raised in both data and theory for the region $25<E_{T}^{\rm jet1}<35$~GeV; the results are 
shown in Fig.~\ref{fig:et2-cut}. With increasing $E_{T}^{\rm jet2, cut}$, the data fall, as 
expected; the trend is well reproduced by the {\sc Herwig} simulation, which includes 
leading-logarithmic parton showers. The prediction from {\sc Herwig} is normalised to the 
data in the first bin in Fig.~\ref{fig:et2-cut}(a). The LO prediction (not shown) for this 
cross section is flat, since only two partons are emitted, which must have equal transverse 
energies by conservation of energy. The predictions of the shape of this distribution from 
$\mathcal{O}$($\alpha \alpha_s^2$) QCD are therefore the lowest non-trivial order 
predictions. The predictions from NLO QCD are shown; they fall less rapidly at low 
$E_{T}^{\rm jet2, cut}$ and more rapidly at high $E_{T}^{\rm jet2, cut}$ than the data. 
Frixione and Ridolfi~\cite{np:b507:315} have shown that, when the requirements on the 
minimum transverse energy of the two jets are similar, the NLO calculation is infrared 
sensitive. This has been investigated further by considering the cross section for regions 
of high and low \xgo. For $\xgo~>~0.8$, shown in Fig.~\ref{fig:et2-cut}(b), the data and 
NLO QCD converge for low $E_{T}^{\rm jet2, cut}$, both being reasonably insensitive to the 
cut and similar in shape. From Fig.~\ref{fig:et2-cut}, it can be seen that for a cut on the 
first jet of $25<E_{T}^{\rm jet1}<35$~GeV, the cut on the jet of lower transverse energy has 
to be below 21~GeV for the NLO predictions to agree with the data. Figure~\ref{fig:et2-cut}(c) 
shows the region $\xgo<0.8$; the predictions lie below the data at low $E_{T}^{\rm jet2, cut}$, 
but within the theoretical uncertainties. The prediction using AFG-HO is about $(10-15)\%$ 
below that of GRV-HO, but is similar in shape and is therefore just compatible with the data. 

The difference in the behaviour of the data and the calculations in Fig.~\ref{fig:et2-cut} implies 
that there is a significant dependence on $E_{T}^{\rm jet2, cut}$ in the comparisons between the 
measurements in the previous section and NLO QCD. By adjusting  $E_{T}^{\rm jet2, cut}$ separately 
in each $E_{T}^{\rm jet1}$ range, it would be possible to achieve agreement between the NLO 
prediction and the data. However, this seems to be a somewhat arbitrary procedure. 

The agreement with theory at high \xgo \ and high transverse energy, where the dependence on the 
photon structure is small, demonstrates a consistency between these data and the gluon distribution 
in the proton extracted from DIS data. Further discrimination between the current PDFs is currently
 not possible given the large 
uncertainties in the theory at low transverse energies and both the experimental and theoretical 
uncertainties at higher transverse energies. However, the data shown here significantly 
constrain the parton densities in the photon. These constraints would be made more stringent 
with improved higher-order, or resummed, calculations.

\section{Conclusions}
\label{sec-conc}

The dijet cross section in photoproduction has been measured in the kinematic region 
$Q^2<1$ GeV$^2$, $0.2<y<0.85$, $E_{T}^{\rm jet1}>14$ GeV, $E_{T}^{\rm jet2}>11$ GeV and 
\mbox{$-1<\eta^{\rm jet1,2}<2.4$}. In the high-mass region defined by $M^{\rm jj} > 42$~GeV and 
$0.1<\bar{\eta}<1.3$, the dijet angular distribution of the data is well reproduced by the 
NLO predictions, indicating that 
the dynamics of the short-distance process is understood. Over the wider region, the measurements 
are compared with NLO predictions using different parameterisations for the parton densities of the 
photon. The data fall less steeply with increasing transverse energy than do the NLO QCD 
predictions, and show sensitivity to the parton densities of the photon. Neither the AFG-HO nor 
the GRV-HO parameterisation, convoluted with the NLO matrix elements, fully describes all features 
of the data. There is agreement with theory at high \xgo \ and high transverse energy, where 
the dependence on the photon structure is small, which represents a consistency check of the gluon 
distribution in the proton extracted from deep inelastic scattering. The data at low \xgo \ 
significantly constrain the parton densities in the photon; future parameterisations of the photon 
PDFs should take them into account. These constraints would be made more 
stringent were improved higher-order or resummed calculations available.

\section*{Acknowledgements}

The design, construction and installation of the ZEUS detector have been made possible by the 
ingenuity and dedicated efforts of many people from inside DESY and from the home institutes 
who are not listed as authors. Their contributions are acknowledged with great appreciation. The 
experiment was made possible by the inventiveness and the diligent efforts of the HERA machine 
group. The strong support and encouragement of the DESY directorate have been invaluable. We 
thank S.~Frixione and G.~Ridolfi for making their NLO code freely available and 
S.~Frixione, in particular, for answering many questions on how to use the code and on jet 
photoproduction in general. We also thank M.~H.~Seymour and M.~Klasen 
for useful discussions.

\vfill\eject


{
\def\bibname{\Large\bf References}
\def\refname{\Large\bf References}
\pagestyle{plain}
\ifzeusbst
  \bibliographystyle{./BiBTeX/bst/l4z_default}
\fi
\ifzdrftbst
  \bibliographystyle{./BiBTeX/bst/l4z_draft}
\fi
\ifzbstepj
  \bibliographystyle{./BiBTeX/bst/l4z_epj}
\fi
\ifzbstnp
  \bibliographystyle{./BiBTeX/bst/l4z_np}
\fi
\ifzbstpl
  \bibliographystyle{./BiBTeX/bst/l4z_pl}
\fi
{\raggedright
\bibliography{./BiBTeX/user/syn.bib,%
              ./BiBTeX/bib/l4z_articles.bib,%
              ./BiBTeX/bib/l4z_books.bib,%
              ./BiBTeX/bib/l4z_conferences.bib,%
              ./BiBTeX/bib/l4z_h1.bib,%
              ./BiBTeX/bib/l4z_misc.bib,%
              ./BiBTeX/bib/l4z_old.bib,%
              ./BiBTeX/bib/l4z_preprints.bib,%
              ./BiBTeX/bib/l4z_replaced.bib,%
              ./BiBTeX/bib/l4z_temporary.bib,%
              ./BiBTeX/bib/l4z_zeus.bib}}
}
\vfill\eject

\renewcommand{\arraystretch}{1.2}
\begin{table}[hbt]
  \begin{center}
  \begin{tabular}{|c|ccccc||c|} \hline 
   $|\cts|$ bin &  $d\sigma/d|\cts|$ & $\Delta_{\rm stat}$ & $\Delta_{\rm syst}$ & $\Delta_{\rm ES}$ & (pb) & $C_{\rm had}$ \\ \hline \hline
   \multicolumn{7}{|c|}{$\xgo < 0.75$} \\ \hline
   0.0, 0.1 &  44.5 & $\pm  3.5$ & $^{+ 1.6}_{- 3.8}$  & $_{ -2.0}^{+ 2.7}$ & & 0.994 $\pm$ 0.010 \\ 
   0.1, 0.2 &  40.1 & $\pm  3.3$ & $^{+ 3.2}_{- 2.4}$  & $_{ -1.3}^{+ 2.3}$ & & 1.027 $\pm$ 0.021 \\ 
   0.2, 0.3 &  48.1 & $\pm  3.6$ & $^{+ 2.5}_{- 3.3}$  & $_{ -2.1}^{+ 2.4}$ & & 1.006 $\pm$ 0.001 \\ 
   0.3, 0.4 &  54.5 & $\pm  3.7$ & $^{+ 4.4}_{- 0.9}$  & $_{ -2.5}^{+ 2.0}$ & & 1.022 $\pm$ 0.017 \\ 
   0.4, 0.5 &  83.9 & $\pm  4.7$ & $^{+ 2.7}_{- 5.4}$  & $_{ -3.6}^{+ 4.5}$ & & 1.008 $\pm$ 0.014 \\ 
   0.5, 0.6 & 115.7 & $\pm  5.5$ & $^{+ 1.4}_{- 4.1}$  & $_{ -4.5}^{+ 4.8}$ & & 1.028 $\pm$ 0.020 \\ 
   0.6, 0.7 & 205.5 & $\pm  7.6$ & $^{+14.7}_{- 4.7}$  & $_{-10.7}^{+ 8.1}$ & & 1.024 $\pm$ 0.010 \\ 
   0.7, 0.8 & 406.9 & $\pm 11.5$ & $^{+ 9.5}_{-32.7}$  & $_{-15.3}^{+16.3}$ & & 1.031 $\pm$ 0.001 \\ \hline
   \multicolumn{7}{|c|}{$\xgo > 0.75$} \\ \hline
   0.0, 0.1 & 141.8 & $\pm  7.1$ & $^{+ 4.0}_{- 8.1}$ & $_{ -5.2}^{+ 5.5}$ & &  0.973 $\pm$ 0.003 \\ 
   0.1, 0.2 & 142.1 & $\pm  7.1$ & $^{+ 2.7}_{-10.0}$ & $_{ -5.6}^{+ 5.7}$ & &  0.969 $\pm$ 0.010 \\ 
   0.2, 0.3 & 155.3 & $\pm  7.3$ & $^{+ 7.9}_{- 6.8}$ & $_{ -6.4}^{+ 6.5}$ & &  0.958 $\pm$ 0.016 \\ 
   0.3, 0.4 & 182.7 & $\pm  7.9$ & $^{+ 5.2}_{- 7.7}$ & $_{ -6.5}^{+ 7.3}$ & &  0.970 $\pm$ 0.007 \\ 
   0.4, 0.5 & 233.8 & $\pm  8.8$ & $^{+ 9.6}_{-10.7}$ & $_{ -8.9}^{+ 8.8}$ & &  0.964 $\pm$ 0.006 \\ 
   0.5, 0.6 & 283.9 & $\pm  9.7$ & $^{+ 9.1}_{-10.8}$ & $_{ -9.9}^{+ 9.4}$ & &  0.963 $\pm$ 0.008 \\ 
   0.6, 0.7 & 439.7 & $\pm 12.2$ & $^{+ 9.2}_{-17.2}$ & $_{-17.8}^{+16.3}$ & &  0.946 $\pm$ 0.020 \\ 
   0.7, 0.8 & 686.6 & $\pm 15.9$ & $^{+26.0}_{-18.3}$ & $_{-22.6}^{+24.8}$ & &  0.933 $\pm$ 0.020 \\ \hline
  \end{tabular}
\end{center}
\caption[]
{Measured cross sections as a function of $|\cts|$ for \xgo~$<0.75$ and \xgo~$>0.75$. The 
statistical, systematic and jet energy scale, $\Delta_{\rm ES}$, uncertainties are shown 
separately. The multiplicative 
hadronisation correction applied to the NLO prediction is shown in the last column. The uncertainty 
shown for the hadronisation correction is half the spread between the values obtained using the 
{\sc Herwig} and {\sc Pythia} models.}
\label{table-cos}
\end{table}

\newpage

\begin{center}
  \begin{tabular}{|c|ccccc|c|} \hline 
   $E_{T}^{\rm jet1}$ bin (GeV) &  $d\sigma/dE_{T}^{\rm jet1}$ & $\Delta_{\rm stat}$ & $\Delta_{\rm syst}$ & $\Delta_{\rm ES}$ & (pb/GeV) & $C_{\rm had}$ \\ \hline \hline
   \multicolumn{7}{|c|}{$-1 < \eta^{\rm jet1,2} < 0$} \\ \hline
   14, 17 & 12.2  & $\pm 0.6$  & $^{+3.6}_{-  1.1}$  & $_{-0.5}^{+ 0.6}$  & & 0.794 $\pm$ 0.034 \\ 
   17, 21 &  3.0  & $\pm 0.3$  & $^{+0.4 }_{- 0.6}$  & $_{-0.1}^{+ 0.1}$  & & 0.757 $\pm$ 0.046 \\ \hline
   \multicolumn{7}{|c|}{$0 < \eta^{\rm jet1} < 1$, $-1 < \eta^{\rm jet2} < 0$} \\ \hline
   14, 17 & 34.3  & $\pm 0.7$  & $^{+2.2}_{-  1.6}$  & $_{-1.0}^{+ 1.3}$  & & 0.909 $\pm$ 0.024 \\ 
   17, 21 & 16.0  & $\pm 0.4$  & $^{+0.5}_{-  2.0}$  & $_{-0.5}^{+ 0.7}$  & & 0.907 $\pm$ 0.016 \\ 
   21, 25 &  4.0  & $\pm 0.2$  & $^{+0.1}_{-  0.5}$  & $_{-0.2}^{+ 0.1}$  & & 0.863 $\pm$ 0.007 \\ 
   25, 29 &  0.74  & $\pm 0.08$  & $^{+0.08}_{-  0.17}$  & $_{-0.03}^{+ 0.02}$  & & 0.834 $\pm$ 0.013 \\ \hline
   \multicolumn{7}{|c|}{$0 < \eta^{\rm jet1,2} < 1$} \\ \hline
   14, 17 & 51.4  & $\pm 1.1$  & $^{+0.4}_{-  5.7}$  & $_{-0.9}^{+ 1.6}$  & & 0.954 $\pm$ 0.030 \\ 
   17, 21 & 26.1  & $\pm 0.7$  & $^{+0.7}_{-  2.1}$  & $_{-1.0}^{+ 0.9}$  & & 0.974 $\pm$ 0.017 \\ 
   21, 25 & 12.3  & $\pm 0.5$  & $^{+0.3}_{-  1.5}$  & $_{-0.4}^{+ 0.4}$  & & 0.973 $\pm$ 0.015 \\ 
   25, 29 &  6.3  & $\pm 0.3$  & $^{+0.1}_{-  0.8}$  & $_{-0.2}^{+ 0.3}$  & & 0.958 $\pm$ 0.021 \\ 
   29, 35 &  1.7  & $\pm 0.1$  & $^{+0.03}_{- 0.2}$  & $_{-0.1}^{+ 0.1}$  & & 0.953 $\pm$ 0.003 \\ 
   35, 41 &  0.56  & $\pm 0.08$  & $^{+0.02}_{- 0.08}$  & $_{-0.02}^{+0.03}$ & & 0.920 $\pm$ 0.006 \\ 
   41, 48 &  0.122  & $\pm 0.037$ & $^{+0.031}_{- 0.014}$ & $_{-0.010}^{+0.003}$ & & 0.872 $\pm$ 0.044 \\ \hline
   \multicolumn{7}{|c|}{$1 < \eta^{\rm jet1} < 2.4$, $-1 < \eta^{\rm jet2} < 0$} \\ \hline
   14, 17 & 28.1  & $\pm 0.6$  & $^{+3.4}_{-  0.7}$  & $_{-1.1}^{+ 1.0}$  & & 0.913 $\pm$ 0.015 \\ 
   17, 21 & 15.8  & $\pm 0.4$  & $^{+0.3}_{-  1.2}$  & $_{-0.6}^{+ 0.6}$  & & 0.922 $\pm$ 0.028 \\ 
   21, 25 &  5.4  & $\pm 0.2$  & $^{+0.2}_{-  0.1}$  & $_{-0.1}^{+ 0.2}$  & & 0.922 $\pm$ 0.012 \\ 
   25, 29 &  1.6  & $\pm 0.1$  & $^{+0.2 }_{- 0.1}$ & $_{-0.1}^{+0.1}$  & & 0.923 $\pm$ 0.012 \\ 
   29, 35 &  0.41  & $\pm 0.05$ & $^{+0.02}_{- 0.05}$ & $_{-0.02}^{+0.01}$ & & 0.862 $\pm$ 0.014 \\ 
   35, 41 &  0.043 & $\pm 0.014$ & $^{+0.013}_{- 0.013}$ & $_{-0.003}^{+0.004}$ & & 0.827 $\pm$ 0.004 \\ \hline
  \end{tabular}
\end{center}

\vspace{0.5cm}
{\bf Table 2:} \emph{Measured cross section as a function of $E_{T}^{\rm jet1}$ for events with $\xgo \ >0.75$. 
The measurement is divided into six regions of the pseudorapidities of the jets. For further details, 
see the caption to Table~\ref{table-cos}.}

\newpage

\begin{table}[hbt]
  \begin{center}
  \begin{tabular}{|c|ccccc|c|} \hline 
   $E_{T}^{\rm jet1}$ bin (GeV) &  $d\sigma/dE_{T}^{\rm jet1}$ & $\Delta_{\rm stat}$ & $\Delta_{\rm syst}$ & $\Delta_{\rm ES}$ & (pb/GeV) & $C_{\rm had}$ \\ \hline \hline
   \multicolumn{7}{|c|}{$1 < \eta^{\rm jet1} < 2.4$, $0 < \eta^{\rm jet2} < 1$} \\ \hline
   14, 17 & 38.9  & $\pm 0.7$  & $^{+0.4}_{-  2.2}$  & $_{-1.1}^{+ 1.1}$  & & 0.990 $\pm$ 0.010 \\ 
   17, 21 & 23.1  & $\pm 0.4$  & $^{+0.2}_{-  1.2}$  & $_{-0.9}^{+ 0.8}$  & & 0.978 $\pm$ 0.011 \\ 
   21, 25 & 11.2  & $\pm 0.3$  & $^{+0.1}_{-  0.6}$  & $_{-0.3}^{+ 0.4}$  & & 0.967 $\pm$ 0.017 \\ 
   25, 29 &  5.3  & $\pm 0.2$  & $^{+0.03}_{- 0.2}$  & $_{-0.2}^{+ 0.2}$  & & 0.969 $\pm$ 0.015 \\ 
   29, 35 &  2.1  & $\pm 0.1$  & $^{+0.01}_{- 0.06}$ & $_{-0.1}^{+ 0.1}$  & & 0.962 $\pm$ 0.022 \\ 
   35, 41 &  0.87  & $\pm 0.07$  & $^{+0.06}_{- 0.05}$ & $_{-0.03}^{+0.06}$ & & 0.971 $\pm$ 0.016 \\ 
   41, 48 &  0.34  & $\pm 0.04$ & $^{+0.01}_{- 0.02}$ & $_{-0.02}^{+0.02}$ & & 0.982 $\pm$ 0.009 \\ 
   48, 55 &  0.13  & $\pm 0.03$ & $^{+0.01}_{- 0.03}$ & $_{-0.01}^{+0.01}$ & & 0.954 $\pm$ 0.023 \\ 
   55, 65 &  0.035 & $\pm 0.011$ & $^{+0.000}_{- 0.014}$ & $_{-0.002}^{+0.002}$ & & 0.974 $\pm$ 0.032 \\ \hline
   \multicolumn{7}{|c|}{$1 < \eta^{\rm jet1,2} < 2.4$} \\ \hline
   14, 17 & 1.5   & $\pm 0.2$  & $^{+0.4 }_{- 0.4}$  & $_{-0.1}^{+ 0.0}$  & & 1.302 $\pm$ 0.225 \\ 
   17, 21 & 4.0   & $\pm 0.2$  & $^{+0.8}_{-  0.5}$  & $_{-0.2}^{+ 0.2}$  & & 1.190 $\pm$ 0.032 \\ 
   21, 25 & 5.0   & $\pm 0.3$  & $^{+0.3 }_{- 0.1}$  & $_{-0.2}^{+ 0.2}$  & & 1.083 $\pm$ 0.032 \\ 
   25, 29 & 3.3   & $\pm 0.2$  & $^{+0.3 }_{- 0.2}$  & $_{-0.1}^{+ 0.1}$  & & 1.017 $\pm$ 0.030 \\ 
   29, 35 & 1.9   & $\pm 0.1$  & $^{+0.1 }_{- 0.1}$  & $_{-0.1}^{+ 0.0}$  & & 1.010 $\pm$ 0.012 \\ 
   35, 41 & 0.93   & $\pm 0.10$  & $^{+0.04 }_{-0.03}$ & $_{-0.02}^{+0.08}$ & & 0.992 $\pm$ 0.009 \\ 
   41, 48 & 0.41   & $\pm 0.06$  & $^{+0.06}_{- 0.02}$ & $_{-0.03}^{+0.02}$ & & 0.982 $\pm$ 0.018 \\ 
   48, 55 & 0.23   & $\pm 0.05$  & $^{+0.03}_{- 0.04}$ & $_{-0.01}^{+0.01}$ & & 1.000 $\pm$ 0.018 \\ 
   55, 65 & 0.033  & $\pm 0.015$ & $^{+0.029}_{- 0.008}$ & $_{-0.002}^{+0.002}$ & & 0.999 $\pm$ 0.004 \\ 
   65, 75 & 0.041  & $\pm 0.017$ & $^{+0.004}_{-0.010}$ & $_{-0.002}^{+0.002}$ & & 0.987 $\pm$ 0.014 \\ \hline
  \end{tabular}
\end{center}
\caption[]
{(cont.)}
\label{table-etgt}
\end{table}

\newpage

\begin{center}
  \begin{tabular}{|c|ccccc|c|} \hline 
   $E_{T}^{\rm jet1}$ bin (GeV) &  $d\sigma/dE_{T}^{\rm jet1}$ & $\Delta_{\rm stat}$ & $\Delta_{\rm syst}$ & $\Delta_{\rm ES}$ & (pb/GeV) & $C_{\rm had}$ \\ \hline \hline
   \multicolumn{7}{|c|}{$-1 < \eta^{\rm jet1,2} < 0$} \\ \hline
   14, 17 & 0.30   & $\pm 0.08$  & $^{+0.07}_{- 0.04}$ & $_{-0.01}^{+0.01}$ & & 1.116 $\pm$ 0.130 \\ \hline
   \multicolumn{7}{|c|}{$0 < \eta^{\rm jet1} < 1$, $-1 < \eta^{\rm jet2} < 0$} \\ \hline
   14, 17 & 6.2   & $\pm 0.3$  & $^{+0.9}_{- 0.3}$  & $_{-0.1}^{+ 0.2}$  & & 1.006 $\pm$ 0.003 \\ 
   17, 21 & 2.1   & $\pm 0.1$  & $^{+0.1}_{- 0.4}$  & $_{-0.1}^{+ 0.1}$  & & 1.014 $\pm$ 0.012 \\ 
   21, 25 & 0.23  & $\pm 0.04$ & $^{+0.05}_{-0.06}$ & $_{-0.01}^{+0.01}$ & & 1.013 $\pm$ 0.027 \\ 
   25, 29 & 0.032  & $\pm 0.015$ & $^{+0.019}_{-0.004}$ & $_{-0.003}^{+0.003}$ & & 0.888 $\pm$ 0.001 \\ \hline
   \multicolumn{7}{|c|}{$0 < \eta^{\rm jet1,2} < 1$} \\ \hline
   14, 17 & 29.4  & $\pm 0.8$  & $^{+0.4}_{- 1.9}$  & $_{-1.1}^{+ 1.1}$  & & 1.006 $\pm$ 0.013 \\ 
   17, 21 & 12.7  & $\pm 0.4$  & $^{+0.4}_{- 1.4}$  & $_{-0.5}^{+ 0.5}$  & & 1.006 $\pm$ 0.005 \\ 
   21, 25 &  4.6  & $\pm 0.3$  & $^{+0.2}_{- 0.5}$  & $_{-0.2}^{+ 0.3}$  & & 1.004 $\pm$ 0.014 \\ 
   25, 29 &  1.2 & $\pm 0.1$  & $^{+0.2}_{- 0.1}$  & $_{-0.04}^{+ 0.10}$  & & 0.972 $\pm$ 0.020 \\ 
   29, 35 &  0.21 & $\pm 0.04$ & $^{+0.07}_{-0.02}$ & $_{-0.01}^{+0.02}$ & & 1.043 $\pm$ 0.038 \\ \hline
   \multicolumn{7}{|c|}{$1 < \eta^{\rm jet1} < 2.4$, $-1 < \eta^{\rm jet2} < 0$} \\ \hline
   14, 17 & 15.4  & $\pm 0.4$  & $^{+2.3}_{- 0.5}$  & $_{-0.5}^{+ 0.8}$  & & 1.040 $\pm$ 0.019 \\ 
   17, 21 &  6.0  & $\pm 0.2$  & $^{+0.5}_{- 0.1}$  & $_{-0.3}^{+ 0.3}$  & & 1.039 $\pm$ 0.017 \\ 
   21, 25 &  1.4  & $\pm 0.1 $ & $^{+0.2}_{- 0.1}$  & $_{-0.04}^{+0.05}$ & & 1.007 $\pm$ 0.019 \\ 
   25, 29 &  0.38  & $\pm 0.05$ & $^{+0.03}_{-0.18}$  & $_{-0.03}^{+0.03}$ & & 1.031 $\pm$ 0.017 \\ 
   29, 35 &  0.057 & $\pm 0.015$ & $^{+0.013}_{-0.022}$ & $_{-0.001}^{+0.001}$ & & 0.913 $\pm$ 0.013 \\ \hline
   \multicolumn{7}{|c|}{$1 < \eta^{\rm jet1} < 2.4$, $0 < \eta^{\rm jet2} < 1$} \\ \hline
   14, 17 & 56.7  & $\pm 0.8$  & $^{+1.6}_{- 1.0}$  & $_{-2.1}^{+ 2.1}$  & & 1.007 $\pm$ 0.024 \\ 
   17, 21 & 26.6  & $\pm 0.4$  & $^{+0.5}_{- 0.5}$  & $_{-1.0}^{+ 1.3}$  & & 1.021 $\pm$ 0.018 \\ 
   21, 25 & 11.2  & $\pm 0.3$  & $^{+0.1}_{- 0.7}$  & $_{-0.5}^{+ 0.6}$  & & 1.013 $\pm$ 0.014 \\ 
   25, 29 & 3.7   & $\pm 0.2$  & $^{+0.1}_{- 0.2}$  & $_{-0.2}^{+ 0.1}$  & & 0.999 $\pm$ 0.015 \\ 
   29, 35 & 1.3   & $\pm 0.1$  & $^{+0.05}_{-0.05}$ & $_{-0.1}^{+0.1}$ & & 1.003 $\pm$ 0.004 \\ 
   35, 41 & 0.35   & $\pm 0.04$ & $^{+0.04}_{-0.04}$ & $_{-0.02}^{+0.03}$ & & 1.026 $\pm$ 0.009 \\ 
   41, 48 & 0.064 & $\pm 0.016$ & $^{+0.007}_{-0.009}$ & $_{-0.005}^{+0.003}$ & & 1.001 $\pm$ 0.004 \\ 
   48, 55 & 0.019  & $\pm 0.009$ & $^{+0.014}_{-0.005}$ & $_{-0.001}^{+0.001}$ & & 1.000 $\pm$ 0.014 \\ \hline
  \end{tabular}
\end{center}

\vspace{0.25cm}
{\bf Table 3:} \emph{Measured cross section as a function of $E_{T}^{\rm jet1}$ for events with $\xgo \ < 0.75$. The 
measurement is divided into six regions of the pseudorapidities of the jets. For further details, 
see the caption to Table~\ref{table-cos}.}

\newpage

\begin{table}[hbt]
\begin{center}
  \begin{tabular}{|c|ccccc|c|} \hline 
   $E_{T}^{\rm jet1}$ bin (GeV) &  $d\sigma/dE_{T}^{\rm jet1}$ & $\Delta_{\rm stat}$ & $\Delta_{\rm syst}$ & $\Delta_{\rm ES}$ & (pb/GeV) & $C_{\rm had}$ \\ \hline \hline
   \multicolumn{7}{|c|}{$1 < \eta^{\rm jet1,2} < 2.4$} \\ \hline
   14, 17 & 78.4  & $\pm 1.4$  & $^{+6.6}_{- 5.4}$  & $_{-2.8}^{+ 2.3}$  & & 1.000 $\pm$ 0.024 \\ 
   17, 21 & 38.2  & $\pm 0.7$  & $^{+2.5}_{- 0.9}$  & $_{-1.6}^{+ 1.7}$  & & 1.019 $\pm$ 0.025 \\ 
   21, 25 & 16.9  & $\pm 0.5$  & $^{+0.9}_{- 0.3}$  & $_{-0.7}^{+ 0.8}$  & & 1.016 $\pm$ 0.025 \\ 
   25, 29 & 6.3   & $\pm 0.3$  & $^{+0.4}_{- 0.1}$  & $_{-0.3}^{+ 0.3}$  & & 1.012 $\pm$ 0.025 \\ 
   29, 35 & 2.5   & $\pm 0.1$  & $^{+0.1}_{- 0.0}$  & $_{-0.1}^{+ 0.1}$  & & 1.020 $\pm$ 0.032 \\ 
   35, 41 & 0.92  & $\pm 0.08$  & $^{+0.05}_{-0.01}$ & $_{-0.06}^{+0.09}$ & & 1.018 $\pm$ 0.031 \\ 
   41, 48 & 0.31  & $\pm 0.05$ & $^{+0.03}_{-0.04}$ & $_{-0.02}^{+0.02}$ & & 1.000 $\pm$ 0.026 \\ 
   48, 55 & 0.12  & $\pm 0.03$ & $^{+0.01}_{-0.01}$ & $_{-0.01}^{+0.01}$ & & 0.995 $\pm$ 0.019 \\ 
   55, 65 & 0.045 & $\pm 0.016$ & $^{+0.014}_{-0.016}$ & $_{-0.003}^{+0.004}$ & & 0.998 $\pm$ 0.003 \\ \hline
  \end{tabular}
\end{center}
\caption[]
{(cont.)}
\label{table-etlt}
\end{table}

\newpage

\begin{table}[hbt]
  \begin{center}
  \begin{tabular}{|c|ccccc|c|} \hline 
   $\eta^{\rm jet2}$ bin &  $d\sigma/d\eta^{\rm jet2}$ & $\Delta_{\rm stat}$ & $\Delta_{\rm syst}$ & $\Delta_{\rm ES}$ & (pb) & $C_{\rm had}$ \\ \hline \hline
   \multicolumn{7}{|c|}{$-1 < \eta^{\rm jet1} < 0$} \\ \hline
   $-$1.0, $-$0.5 &   8.9 & $\pm 1.0$ & $^{+ 9.5}_{- 1.4}$ & $_{-0.8}^{+ 0.5}$  & & 0.538 $\pm$ 0.021 \\ 
   $-$0.5,    0.0 &  88.2 & $\pm 2.8$ & $^{+15.6}_{- 9.2}$ & $_{-3.2}^{+ 4.2}$  & & 0.805 $\pm$ 0.031 \\ 
      0.0,    0.5 & 175.0 & $\pm 3.8$ & $^{+11.1}_{-15.7}$ & $_{-5.7}^{+ 5.8}$  & & 0.881 $\pm$ 0.016 \\ 
      0.5,    1.0 & 195.8 & $\pm 4.0$ & $^{+ 6.7}_{-12.4}$ & $_{-6.3}^{+ 8.5}$  & & 0.906 $\pm$ 0.012 \\ 
      1.0,    1.5 & 160.3 & $\pm 3.4$ & $^{+ 9.8}_{- 5.0}$ & $_{-5.9}^{+ 6.5}$  & & 0.921 $\pm$ 0.012 \\ 
      1.5,    2.0 & 130.0 & $\pm 3.3$ & $^{+ 6.9}_{- 2.6}$ & $_{-4.7}^{+ 4.2}$  & & 0.913 $\pm$ 0.021 \\ 
      2.0,    2.4 &  89.3 & $\pm 3.5$ & $^{+ 5.1}_{- 5.6}$ & $_{-3.1}^{+ 3.6}$  & & 0.899 $\pm$ 0.021 \\ \hline
   \multicolumn{7}{|c|}{$0 < \eta^{\rm jet1} < 1$} \\ \hline
   $-$1.0, $-$0.5 &  99.3 & $\pm 3.1$ & $^{+11.1}_{- 1.9}$ & $_{- 2.8}^{+ 4.7}$ & & 0.805 $\pm$ 0.012 \\ 
   $-$0.5,    0.0 & 272.0 & $\pm 4.6$ & $^{+10.7}_{-26.6}$ & $_{- 9.1}^{+ 9.8}$ & & 0.922 $\pm$ 0.010 \\ 
      0.0,    0.5 & 347.0 & $\pm 5.1$ & $^{+ 3.6}_{-37.8}$ & $_{-10.7}^{+12.1}$ & & 0.952 $\pm$ 0.015 \\ 
      0.5,    1.0 & 347.1 & $\pm 5.0$ & $^{+ 2.2}_{-32.5}$ & $_{- 9.2}^{+11.7}$ & & 0.960 $\pm$ 0.013 \\ 
      1.0,    1.5 & 287.9 & $\pm 4.2$ & $^{+ 2.7}_{-19.5}$ & $_{- 9.7}^{+ 9.3}$ & & 0.971 $\pm$ 0.004 \\ 
      1.5,    2.0 & 208.1 & $\pm 3.7$ & $^{+ 1.6}_{- 5.4}$ & $_{- 7.2}^{+ 7.5}$ & & 0.978 $\pm$ 0.012 \\ 
      2.0,    2.4 & 125.1 & $\pm 3.7$ & $^{+ 1.9}_{- 7.9}$ & $_{- 4.3}^{+ 3.9}$ & & 0.967 $\pm$ 0.020 \\ \hline
   \multicolumn{7}{|c|}{$1 < \eta^{\rm jet1} < 2.4$} \\ \hline
   $-$1.0, $-$0.5 & 115.4 & $\pm 3.2$ & $^{+20.5}_{- 6.8}$ & $_{- 4.8}^{+ 5.2}$ & & 0.866 $\pm$ 0.022 \\ 
   $-$0.5,    0.0 & 241.7 & $\pm 4.2$ & $^{+ 6.6}_{- 3.5}$ & $_{- 8.3}^{+ 8.5}$ & & 0.933 $\pm$ 0.011 \\ 
      0.0,    0.5 & 315.7 & $\pm 4.6$ & $^{+ 3.8}_{-13.3}$ & $_{-11.2}^{+11.4}$ & & 0.953 $\pm$ 0.018 \\ 
      0.5,    1.0 & 277.5 & $\pm 4.2$ & $^{+ 1.2}_{-18.6}$ & $_{- 9.1}^{+ 8.5}$ & & 0.992 $\pm$ 0.002 \\ 
      1.0,    1.5 & 104.0 & $\pm 2.4$ & $^{+ 4.1}_{- 3.5}$ & $_{- 4.0}^{+ 3.8}$ & & 1.021 $\pm$ 0.007 \\ 
      1.5,    2.0 &  37.2 & $\pm 1.5$ & $^{+ 0.8}_{- 1.4}$ & $_{- 1.3}^{+ 1.5}$ & & 1.018 $\pm$ 0.008 \\ 
      2.0,    2.4 &  13.2 & $\pm 1.1$ & $^{+ 0.9}_{- 2.3}$ & $_{- 0.4}^{+ 0.5}$ & & 1.031 $\pm$ 0.017 \\ \hline
  \end{tabular}
\end{center}
\caption[]
{Measured cross section as a function of $\eta^{\rm jet2}$ for events with $\xgo \ > 0.75$. The 
measurement is divided into three regions of the pseudorapidity of the other 
jet. For further details, see the caption to Table~\ref{table-cos}.}
\label{table-etagt}
\end{table}

\newpage

\begin{table}[hbt]
  \begin{center}
  \begin{tabular}{|c|ccccc|c|} \hline 
   $\eta^{\rm jet2}$ bin &  $d\sigma/d\eta^{\rm jet2}$ & $\Delta_{\rm stat}$ & $\Delta_{\rm syst}$ & $\Delta_{\rm ES}$ & (pb) & $C_{\rm had}$ \\ \hline \hline
   \multicolumn{7}{|c|}{$-1 < \eta^{\rm jet1} < 0$} \\ \hline
   $-$0.5,    0.0 &  2.2  & $\pm 0.4$ & $^{+1.4}_{-0.8}$   & $_{-0.1}^{+0.1}$   & & 1.088 $\pm$ 0.113 \\ 
      0.0,    0.5 & 15.8  & $\pm 1.0$ & $^{+2.9}_{-0.5}$   & $_{-0.6}^{+0.6}$   & & 1.003 $\pm$ 0.019 \\ 
      0.5,    1.0 & 40.0  & $\pm 1.7$ & $^{+2.7}_{-2.2}$   & $_{-1.0}^{+1.3}$   & & 1.008 $\pm$ 0.008 \\ 
      1.0,    1.5 & 52.4  & $\pm 1.8$ & $^{+6.2}_{-0.4}$   & $_{-1.8}^{+2.6}$   & & 1.033 $\pm$ 0.027 \\ 
      1.5,    2.0 & 60.4  & $\pm 2.0$ & $^{+7.5}_{-4.6}$   & $_{-2.7}^{+3.0}$   & & 1.028 $\pm$ 0.002 \\ 
      2.0,    2.4 & 54.1  & $\pm 2.4$ & $^{+6.9}_{-3.5}$   & $_{-1.4}^{+2.7}$   & & 1.038 $\pm$ 0.024 \\ \hline
   \multicolumn{7}{|c|}{$0 < \eta^{\rm jet1} < 1$} \\ \hline
   $-$1.0, $-$0.5 &   4.1 & $\pm 0.5$ & $^{+1.5}_{- 0.6}$  & $_{ -0.1}^{+ 0.3}$ & & 1.046 $\pm$ 0.020 \\ 
   $-$0.5,    0.0 &  51.7 & $\pm 1.9$ & $^{+4.7}_{- 2.2}$  & $_{ -1.6}^{+ 1.6}$ & & 1.004 $\pm$ 0.002 \\ 
      0.0,    0.5 & 131.7 & $\pm 3.1$ & $^{+3.3}_{-11.7}$  & $_{ -5.2}^{+ 6.6}$ & & 1.002 $\pm$ 0.005 \\ 
      0.5,    1.0 & 195.4 & $\pm 3.8$ & $^{+4.4}_{-14.4}$  & $_{ -7.7}^{+ 7.4}$ & & 1.002 $\pm$ 0.002 \\ 
      1.0,    1.5 & 230.1 & $\pm 3.7$ & $^{+4.8}_{-11.3}$  & $_{ -9.4}^{+ 9.7}$ & & 0.998 $\pm$ 0.023 \\ 
      1.5,    2.0 & 262.4 & $\pm 4.1$ & $^{+5.7}_{- 4.1}$  & $_{-10.2}^{+11.5}$ & & 1.012 $\pm$ 0.014 \\ 
      2.0,    2.4 & 243.7 & $\pm 5.0$ & $^{+6.7}_{- 7.1}$  & $_{ -9.1}^{+10.1}$ & & 1.020 $\pm$ 0.011 \\ \hline
   \multicolumn{7}{|c|}{$1 < \eta^{\rm jet1} < 2.4$} \\ \hline
   $-$1.0, $-$0.5 &  20.5 & $\pm 1.2$ & $^{+ 5.1}_{- 0.3}$ & $_{ -0.8}^{+ 0.9}$ & & 1.056 $\pm$ 0.020 \\ 
   $-$0.5,    0.0 & 135.0 & $\pm 3.0$ & $^{+13.5}_{- 5.1}$ & $_{ -5.0}^{+ 6.9}$ & & 1.028 $\pm$ 0.022 \\ 
      0.0,    0.5 & 288.8 & $\pm 4.4$ & $^{+ 5.6}_{- 6.5}$ & $_{-11.6}^{+12.1}$ & & 1.016 $\pm$ 0.013 \\ 
      0.5,    1.0 & 403.6 & $\pm 5.2$ & $^{+ 9.6}_{- 9.7}$ & $_{-15.5}^{+17.5}$ & & 1.006 $\pm$ 0.018 \\ 
      1.0,    1.5 & 403.0 & $\pm 4.8$ & $^{+24.2}_{-15.6}$ & $_{-16.0}^{+16.3}$ & & 1.008 $\pm$ 0.029 \\ 
      1.5,    2.0 & 355.2 & $\pm 4.5$ & $^{+27.6}_{-20.8}$ & $_{-14.4}^{+13.6}$ & & 1.013 $\pm$ 0.026 \\ 
      2.0,    2.4 & 305.0 & $\pm 5.3$ & $^{+27.4}_{-11.1}$ & $_{-10.8}^{+11.0}$ & & 1.015 $\pm$ 0.021 \\ \hline
  \end{tabular}
\end{center}
\caption[]
{Measured cross section as a function of $\eta^{\rm jet2}$ for events with $\xgo \ < 0.75$. The 
measurement is divided into three regions of the pseudorapidity of the other 
jet. For further details, see the caption to Table~\ref{table-cos}.}
\label{table-etalt}
\end{table}

\newpage

\begin{table}[hbt]
  \begin{center}
  \begin{tabular}{|c|ccccc|c|} \hline 
   \xgo \ bin &  $d\sigma/d\xgo$ & $\Delta_{\rm stat}$ & $\Delta_{\rm syst}$ & $\Delta_{\rm ES}$ & (pb) & $C_{\rm had}$ \\ \hline \hline
   \multicolumn{7}{|c|}{$14 < E_{T}^{\rm jet1} < 17$ GeV} \\ \hline
   0.1, 0.2 &  421.4 & $\pm 14.1$ & $^{+49.9}_{-33.5}$  & $_{-16.6}^{+19.5}$ & & 0.970 $\pm$ 0.002 \\ 
   0.2, 0.3 &  631.6 & $\pm 16.2$ & $^{+48.2}_{-56.9}$  & $_{-25.7}^{+19.3}$ & & 0.994 $\pm$ 0.027 \\ 
   0.3, 0.4 &  607.2 & $\pm 15.0$ & $^{+38.2}_{-14.3}$  & $_{-23.0}^{+20.2}$ & & 0.992 $\pm$ 0.022 \\ 
   0.4, 0.5 &  599.9 & $\pm 14.4$ & $^{+36.7}_{-4.2}$   & $_{-19.9}^{+26.6}$ & & 1.002 $\pm$ 0.023 \\ 
   0.5, 0.6 &  621.2 & $\pm 14.6$ & $^{+49.8}_{-6.5}$   & $_{-17.9}^{+21.6}$ & & 1.016 $\pm$ 0.026 \\ 
   0.6, 0.7 &  695.8 & $\pm 15.4$ & $^{+33.2}_{-12.0}$  & $_{-24.5}^{+24.6}$ & & 1.035 $\pm$ 0.014 \\ 
   0.7, 0.8 &  842.1 & $\pm 16.5$ & $^{+97.5}_{-6.6}$   & $_{-25.9}^{+33.4}$ & & 1.113 $\pm$ 0.002 \\ 
   0.8, 1.0 & 1784.0 & $\pm 18.8$ & $^{+128.6}_{-78.1}$ & $_{-52.7}^{+57.5}$ & & 0.906 $\pm$ 0.016 \\ \hline
   \multicolumn{7}{|c|}{$17 < E_{T}^{\rm jet1} < 25$ GeV} \\ \hline
   0.1, 0.2 &  251.2 & $\pm  9.6$ & $^{+11.2}_{-15.0}$  & $_{-13.3}^{+15.1}$ & & 1.006 $\pm$ 0.019 \\ 
   0.2, 0.3 &  419.2 & $\pm 11.3$ & $^{+28.5}_{-20.3}$  & $_{-19.9}^{+20.3}$ & & 1.005 $\pm$ 0.013 \\ 
   0.3, 0.4 &  523.1 & $\pm 12.5$ & $^{+39.1}_{-21.3}$  & $_{-21.2}^{+27.3}$ & & 1.008 $\pm$ 0.019 \\ 
   0.4, 0.5 &  539.7 & $\pm 12.5$ & $^{+10.3}_{-14.3}$  & $_{-23.2}^{+22.6}$ & & 1.008 $\pm$ 0.012 \\ 
   0.5, 0.6 &  597.3 & $\pm 13.2$ & $^{+15.3}_{-18.1}$  & $_{-24.7}^{+28.0}$ & & 1.025 $\pm$ 0.018 \\ 
   0.6, 0.7 &  641.3 & $\pm 13.6$ & $^{+21.3}_{-12.6}$  & $_{-23.1}^{+29.4}$ & & 1.022 $\pm$ 0.019 \\ 
   0.7, 0.8 &  812.2 & $\pm 15.2$ & $^{+21.2}_{-21.0}$  & $_{-32.3}^{+34.0}$ & & 1.079 $\pm$ 0.002 \\ 
   0.8, 1.0 & 1791.5 & $\pm 17.9$ & $^{+30.6}_{-154.1}$ & $_{-62.4}^{+65.9}$ & & 0.933 $\pm$ 0.018 \\ \hline
   \multicolumn{7}{|c|}{$25 < E_{T}^{\rm jet1} < 35$ GeV} \\ \hline
   0.1, 0.4 &  40.0  & $\pm 1.8$  & $^{+8.4}_{-0.7}$    & $_{-2.2}^{+2.0}$   & & 1.002 $\pm$ 0.025 \\ 
   0.4, 0.6 &  89.8  & $\pm 3.3$  & $^{+5.4}_{-2.1 }$   & $_{-3.5}^{+3.9}$   & & 1.011 $\pm$ 0.011 \\ 
   0.6, 0.8 & 130.0  & $\pm 4.0$  & $^{+2.3}_{-3.3}$    & $_{-5.7}^{+5.1}$   & & 1.019 $\pm$ 0.021 \\ 
   0.8, 1.0 & 332.2  & $\pm 7.3$  & $^{+30.7}_{-32.0}$  & $_{-12.7}^{+11.6}$ & & 0.948 $\pm$ 0.016 \\ \hline
   \multicolumn{7}{|c|}{$35 < E_{T}^{\rm jet1} < 90$ GeV} \\ \hline
   0.1, 0.4 &  3.9   & $\pm 0.6$  & $^{+0.4}_{-2.0}$   & $_{-0.2}^{+0.3}$    & & 0.991 $\pm$ 0.018 \\ 
   0.4, 0.6 & 12.7   & $\pm 1.2$  & $^{+1.1}_{-0.3}$   & $_{-0.8}^{+1.1}$    & & 1.017 $\pm$ 0.020 \\ 
   0.6, 0.8 & 24.5   & $\pm 1.7$  & $^{+1.3}_{-2.5}$   & $_{-1.3}^{+2.2}$    & & 1.017 $\pm$ 0.015 \\ 
   0.8, 1.0 & 77.6   & $\pm 3.6$  & $^{+1.6}_{-6.1}$   & $_{-3.3}^{+4.5}$    & & 0.961 $\pm$ 0.011 \\ \hline
  \end{tabular}
\end{center}
\caption[]
{Measured cross section as a function of \xgo \ in four regions of $E_T^{\rm jet1}$. 
For further details, see the caption to Table~\ref{table-cos}.}
\label{table-xgamma}
\end{table}


%
%
\end{document}